\documentclass[journal, twocolumn, 10pt]{IEEEtran}
\usepackage[justification=centering]{caption}
\usepackage[margin=1.75cm,nohead]{geometry}

\ifCLASSINFOpdf
   \usepackage[pdftex]{graphicx}
\else
\fi

\usepackage[cmex10]{amsmath}

\usepackage{subfigure}
\usepackage{url}
\usepackage{color}

\usepackage{setspace}

\begin{document}

\title{Flow Allocation for Maximum Throughput and Bounded Delay on Multiple Disjoint Paths for Random Access Wireless Multihop Networks}

\author{Manolis Ploumidis, Nikolaos Pappas, Apostolos Traganitis
\thanks{This work was presented in part in 9th IEEE Broadband Wireless Access Workshop \cite{BWA}.}
\thanks{M.~Ploumidis and A.~Traganitis are with the
Institute of Computer Science, Foundation for Research and Technology - Hellas (FORTH)
and the Computer Science Department, University of Crete, Greece email:\{ploumid,tragani\}@ics.forth.gr}
\thanks{N.~Pappas is with the Department of Science and Technology, Link\"{o}ping University, Norrk\"{o}ping SE-60174, Sweden email:nikolaos.pappas@liu.se}
\thanks{M. Ploumidis was supported by ``HERACLEITUS II -
University of Crete", NSRF (ESPA) (2007-2013) and was co-funded by the European Union and national resources. The research leading to these results has received funding from the People Programme (Marie Curie Actions) of the European Union's Seventh Framework Programme FP7/2007-2013/ under REA grant agreement n$^o$ [612361] (SOrBet).}}

\maketitle

\begin{abstract}
In this paper, we consider random access, wireless, multi-hop networks, with multi-packet reception capabilities, where multiple
flows are forwarded to the gateways through node disjoint paths.
We explore the issue of allocating flow on multiple paths, exhibiting both intra- and inter-path interference, in order to maximize
average aggregate flow throughput (AAT) and also provide bounded packet delay.
A distributed flow allocation scheme is proposed where allocation of flow on paths is formulated as an optimization problem.
Through an illustrative topology it is shown that the corresponding problem is non-convex.
Furthermore, a simple, but accurate model is employed for the average aggregate throughput achieved by all flows, that captures both intra- and
inter-path interference through the SINR model.
The proposed scheme is evaluated through Ns2 simulations of several random wireless scenarios.
Simulation results reveal that, the model employed, accurately captures the AAT observed in the simulated scenarios, even when
the assumption of saturated queues is removed.
Simulation results also show that the proposed scheme achieves significantly higher AAT, for the vast majority of the wireless scenarios explored,
than the following flow allocation schemes:
one that assigns flows on paths on a round-robin fashion, one that optimally utilizes the best path only, and another one that assigns
the maximum possible flow on each path.
Finally, a variant of the proposed scheme is explored, where interference for each link is approximated by considering its dominant interfering
nodes only.
\end{abstract}


\IEEEpeerreviewmaketitle

\section{Introduction}
In order to better utilize the scarce resources of wireless multi-hop networks and meet the increased user demand for QoS,
numerous studies have suggested the use of multiple paths in parallel.
Utilization of multiple paths can provide a wide range of benefits in terms of, throughput \cite{6133896, b:wowmom2010, b:PloumidisCOMCOM},
delay \cite{mp_route_wim, b:wowmom2010, b:PloumidisCOMCOM}, reliability \cite{mp_route_mpdsr, b:wowmom2010, b:PloumidisCOMCOM},
load balancing \cite{5198989, 6133896}, security \cite{ref_mhrp} and energy efficiency \cite{5198989,5425265}.
However, multipath utilization in wireless networks, is more complicated compared to their wired counterparts, since transmissions across a link
interfere with neighbouring links, reducing thus, network performance.

\subsection{Related work}
\label{sec:related_work}

A wide range of different schemes, have been proposed in literature, focusing on multipath utilization for improving network performance, including
routing schemes, resource allocation, flow control, opportunistic-based forwarding ones, e.t.c.
A significant amount of studies focuses on identifying the set of paths that will guarantee improved performance, in terms of some
metric \cite{6133896, mp_route_wim, mp_route_cop_pkt_cach, mp_route_mplot}.
However, such studies, mostly address the issue of, \textit{which} paths should be utilized and rely on heuristic-based approaches concerning the issue
of \textit{how}, should these paths be utilized.
In \cite{mp_route_cop_pkt_cach} for example, traffic is allocated on a round-robin fashion among the available paths.

Several studies suggest schemes that perform joint scheduling with routing, power control or channel assignment \cite{1717611,5501845,6130552}.
As far as, flow allocation on multiple paths and rate control are concerned, a well studied approach associates a utility function to each flow's rate
and aims at maximizing the sum of these utilities, subject to cross-layer constraints.
Along this direction, several studies suggest, joint congestion control and scheduling approaches \cite{1430253, 4509706, 4146795}.
Authors in \cite{5089987}, instead of employing a utility function of a flow's rate, they employ a utility function of flow's effective
rate, in order to take into account the effect of lossy links.
Different from these approaches, this work considers random access networks and also no scheduling is assumed, or devised.

The utility maximization framework, has also been applied in the context of random access networks for designing joint congestion
and contention control schemes \cite{4432241, 1665008}. As far as the interference model in these studies is concerned, no
capture is assumed
and thus, concurrent transmissions on interfering links fail each other. A joint routing and MAC control scheme, for wireless random access
networks, is explored in \cite{5425726}, where interference is modelled through conflict sets and the SINR model.
Different from all these studies, this work considers wireless random access networks, where interference is captured through
the SINR model, taking also into account the effect of Rayleigh fading on signal attenuation.

Authors in \cite{6565329}, study an MPLS-based forwarding paradigm and aim at identifying a feasible routing solution, for multiple flows deploying multiple
paths. Links whose transmissions have a significant effect on each others success probability, are
considered to belong to the same collision domain and cannot be active at the same time.
In \cite{DiStasi201426}, a technique for combining multipath forwarding with packet aggregation, over IEEE802.11 wireless mesh networks, is suggested.
Multipath utilization is accomplished by employing Layer-2.5, a multipath routing and forwarding strategy, that aims at utilizing links
in proportion to their available bandwidth.
Authors in \cite{1530015}, suggest a distributed rate allocation algorithm, aimed at minimizing the total distortion of video streams,
transmitted over wireless adhoc networks.
In \cite{Huang:2001:MFS:501416.501447}, a max-min fair scheduling allocation algorithm is proposed, along with a modified backoff algorithm, for achieving
long term fairness.
In \cite{1296640}, a distributed flow control algorithm, aimed at maximizing the total traffic flowing from sources to destinations, also
providing network lifetime guarantees, is proposed.
Based on the theoretical ideas of back-pressure scheduling and utility maximization, Horizon \cite{horizon_2008}, constitutes
a practical implementation of a multipath forwarding scheme that interacts with TCP.

There is also a significant amount of studies that suggest
opportunistic forwarding/routing schemes that exploit the broadcast nature of the wireless medium. \cite{Gkantsidis:2007:MCC:1364654.1364667},
suggests a multipath routing protocol called Multipath Code Casting that employs opportunistic forwarding combined with network coding.
It also performs congestion control and employs a rate control mechanism, that achieves fairness among different flows, by maximizing
an aggregate utility of these flows. Authors in \cite{4509888}, suggest an optimization framework that performs optimal flow control, routing,
scheduling and rate adaptation, employing multiple paths and opportunistic transmissions.
Other works consider network-level cooperation combining queueing analysis but focusing on simple topologies \cite{b:PappasTWC2015, b:PappasITW2011, b:PapadimitriouCOMNET}.

\subsection{Contributions}
\label{sec:contr}
Different from all the above, in this study, we consider wireless, random access, multihop networks, with multi-packet reception capabilities,
where multiple unicast flows are forwarded to their destinations through multiple paths that share no common nodes (node disjoint paths).
It should also be noted that, the assumption of random access implies that transmitters get access to the shared medium in a decentralized manner,
without presupposing any coordination method.
We address the problem of allocating flow data rates on paths in such a way that they maximize the average aggregate flow throughput, while also
providing bounded packet delay, in the presence of both intra- and inter-path interference, for the aforementioned type of networks.
For the rest of the paper, we will refer to this problem as the \textit{flow allocation problem}.
The main contribution of this study, is a scheme that formulates flow rate allocation as an optimization problem.
The key feature of this scheme is its distributed nature; the information that is needed to be propagated for each node through the routing
protocol is the position of each node, an indication of whether a node is a flow source, relay, or sink and the transmission
probability in case of a relay node. With this information available, each node can infer its own instance of the topology and also its own
instance of the aforementioned flow allocation problem. Each flow source can then solve this problem, independently of all other sources
in order to derive flow rates that collectively, maximize AAT.
Another key feature of the proposed scheme, is that it maximizes the average aggregate throughput (AAT) achieved by all flows, while also providing
bounded packet delay guarantees.
For the rest of this study we will refer to this scheme as, the \textit{Throughput Optimal Flow Rate Allocation} scheme, or \textit{TOFRA},
for reasons of brevity.
The proposed scheme is based on a simple, but accurate model for the average aggregate throughput, capturing
both inter- and intra-path interference through the SINR model.
Additionally, the effect of Rayleigh fading is also taken into account for deriving a link's success probability.
Through a simple topology we show that the corresponding flow allocation optimization problem is non-convex.
Another contribution of the study, is the evaluation of the proposed scheme, through Ns2 simulations of several random wireless scenarios.

In the evaluation process, the accuracy of the model, for capturing the AAT observed in the simulation scenarios, is explored.
Simulation results show that, the model employed by the proposed scheme, accurately captures the AAT observed in the simulation results, even
when the assumption of saturated queues is removed.
In the second part of the evaluation process, we compare the AAT achieved by the
proposed flow allocation scheme with the following flow allocation schemes: \textit{Best-path}, that optimally utilizes the best path available,
\textit{Full MultiPath}, that
assigns the maximum possible flow (one packet per slot) on each path, and a \textit{Round-Robin} based one. For all simulated scenarios
and all SINR threshold values
considered, the proposed scheme achieves significantly higher throughput than full multipath and Best-path. Additionally, the proposed scheme
outperforms round robin-based flow allocation for the vast majority of the scenarios explored.
Finally, a key contribution of the study is that we also explore a variant of the proposed scheme, where interference for a link
is approximated by considering the \textit{dominant} interfering nodes for that link only.
More precisely, we explore the trade-off between accurately capturing the AAT observed in the simulation scenarios and the complexity
in formulating and solving, the corresponding optimization problem.

The rest of the paper is organized as follows: Section~\ref{sec:system_model}, presents the system model considered.
In Section \ref{sec_tofra_scheme},
we present the proposed flow allocation scheme and demonstrate it
through a simple topology. In Section \ref{sec:evaluation}, we describe the simulation setup and present the evaluation process.
We conclude this study in Section \ref{sec:conclusions}.

\section{System Model}
\label{sec:system_model}

We consider static, wireless, multi-hop networks, with the following properties:
\begin{IEEEitemize}
    \item Random access to the shared medium where each node transmits independently of all other nodes, based on its transmission probability.
    In this way, no coordination among nodes is required thus, random access is a fully distributed access protocol.
    For flow originators, transmission probability denotes the rate at which they inject packets into the network (flow rate).
    For the relay nodes, transmission probability is fixed to a specific value and no control is assumed.
    \item Time is slotted and each packet transmission requires one timeslot.
    \item Flows among different pairs of source and destination nodes, carry unicast traffic of same-sized packets.
    \item All nodes use the same channel and rate, and are equipped with multi-user detectors being thus, able to successfully decode packets
    from more than one transmitter at the same slot \cite{Verdu:1998:MD:521411}.
    \item We assume that all nodes are half-duplex and thus, cannot transmit and receive simultaneously.
    \item We also assume that, all nodes always have packets available for transmission. However, in the evaluation process, we also consider
    the case that the nodes can have empty queues. As illustrated in Section \ref{sec:sim_results}, there is no significant impact on the AAT.
    \item As far as routing is concerned, multiple, node disjoint paths are assumed to be available by the routing protocol, one for each flow.
    Moreover, source routing is assumed, ensuring that packets of the same flow are routed to the destination along the same path.
    Apart from that, for each node, its position, transmission probability, or flow rate, along with an indication of whether it is a flow
    originator, relay, or sink, are assumed known to all other nodes. This information can be periodically propagated throughout the network through a link-state routing protocol.
\end{IEEEitemize}

\begin{figure}[ht]
\centering
\subfigure[Wireless multi-hop mesh network]{
\includegraphics[scale=0.35]{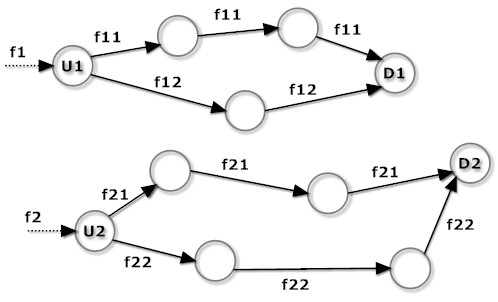}
\label{fig:wireless_scen_1}
}
\subfigure[Sensor network]{
\includegraphics[scale=0.35]{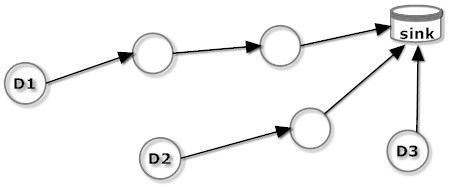}
\label{fig:wireless_scen_2}
}
\label{fig:wireless_scenario}
\caption{Wireless scenarios where throughput optimal flow allocation is applicable.}
\end{figure}

Fig. \ref{fig:wireless_scen_1} and Fig. \ref{fig:wireless_scen_2}, present two different wireless settings, where the suggested flow allocation scheme
may be employed.
In the first scenario, depicted in Fig. \ref{fig:wireless_scen_1}, different users (U1 and U2), generate flows (f1, f2), that are routed to destination nodes
D1 and D2, respectively, through node disjoint multi-hop paths. These flows can be split into multiple subflows, in order to aggregate network resources
and achieve higher aggregate throughput. The suggested flow allocation framework can be applied, in order to identify the data rates for
subflows f11, f12, f21, and f22 that result in maximum average aggregate throughput for both users.

The second scenario, depicted in Fig. \ref{fig:wireless_scen_2}, represents a sensor network, where multiple sensor nodes generate data (D1, D2, and D3),
that are forwarded to the sink, through multiple paths comprised by relay sensor nodes. The proposed flow allocation scheme can be employed
for maximizing the rate at which the sink receives data from the sensor nodes.
\subsection{Physical Layer Model}

\begin{table}[t]
\begin{center}
\begin{tabular}{|c|l|}
\hline
Notation & Definition\\ \hline
$T$ & Set of concurently active transmitters with \\
& link $(i,j)$ \\ \hline
$\alpha$ & Path loss exponent\\ \hline
$\eta_{j}$ & Receiver noise power at $j$\\ \hline
$\gamma_{j}$ & SINR threshold at $j$\\ \hline
$P_{tx}(i)$ & Transmitting power of node $i$\\ \hline
$A(i,j)$ & Random variable for channel fading over \\
& link $(i,j)$\\ \hline
$v(i,j)$ & Prameter of the Rayleigh random variable for \\
& fading over link $(i,j)$\\ \hline
$r(i,j)$ & Distance between nodes $i$, and $j$\\ \hline
$g(i,j)$ & Received power factor for link $(i,j)$\\ \hline
$P_{rx}(i,j)$ & Received power over link $(i,j)$\\ \hline
$p_{i/T}^{j}$ & Success probability for link $(i,j)$, given that \\
& nodes in $T$ are concurrently active\\ \hline
$q_{i}$ & Transmission probability for node $i$, given there \\
& is packet for transmission in its queue\\ \hline
\end{tabular}
\end{center}
\caption{System model related notations}
\label{tab_sys_mod_notations}
\end{table}

The MPR channel model used in this paper is a generalized form of the packet erasure model.
Note that, the notations used for presenting the channel model considered, are also summarized in Table~\ref{tab_sys_mod_notations}.
In the wireless environment, a packet can be decoded correctly by the receiver, if the received $SINR$ exceeds a certain threshold.
More precisely, suppose that a set of nodes, denoted by $T$, is concurrently active with transmitting node $i$, in the same time slot.
Let $P_{rx}(i,j)$ be the signal power received from node $i$ at node $j$.
Let $SINR(i,j)$ be expressed using (\ref{eq:sinr_thres}).
\begin{equation}
\label{eq:sinr_thres}
SINR(i,j)=\frac{P_{rx}(i,j)}{\eta_{j}+\sum_{k\in T\backslash\left\{i\right\}} {P_{rx}(k,j)}}.
\end{equation}
In the above equation, $\eta_{j}$ denotes the receiver noise power at $j$. We assume that a packet transmitted by $i$, is successfully received by $j$,
if and only if, $SINR(i,j)\geq \gamma_{j}$, where $\gamma_{j}$ is a threshold characteristic of node $j$. The wireless channel is subject to fading;
let $P_{tx}(i)$ be the transmitting power of node $i$ and $r(i,j)$ be the distance between $i$ and $j$. The power received by $j$, when $i$
transmits, is $P_{rx}(i,j)=A(i,j)g(i,j)$, where $A(i,j)$ is a random variable representing channel fading. Under Rayleigh fading, it is
known~\cite{b:Tse} that $A(i,j)$ is exponentially distributed. The received power factor, $g(i,j)$, is given by $g(i,j)=P_{tx}(i)(r(i,j))^{-\alpha}$,
where $\alpha$ is the path loss exponent, with typical values between $2$ and $4$. The success probability of link $(i,j)$, when the transmitting nodes
are in $T$, is given by:

\begin{equation}
\label{eq:succprob}
p_{i/T}^{j}=\exp\left(-\frac{\gamma_{j}\eta_{j}}{v(i,j)g(i,j)}\right) \prod_{k\in T\backslash \left\{i,j\right\}}{\left(1+\gamma_{j}\frac{v(k,j)g(k,j)}{v(i,j)g(i,j)}\right)}^{-1},
\end{equation}
where $v(i,j)$ is the parameter of the Rayleigh random variable for fading. The analytical derivation for this success probability,
which captures the effect of interference on link $(i,j)$, from transmissions of nodes in set $T$, can be found in~\cite{b:Nguyen}.

\section{Throughput optimal flow rate allocation (TOFRA) scheme}
\label{sec_tofra_scheme}

\begin{figure}
\centering
\includegraphics[scale=0.3]{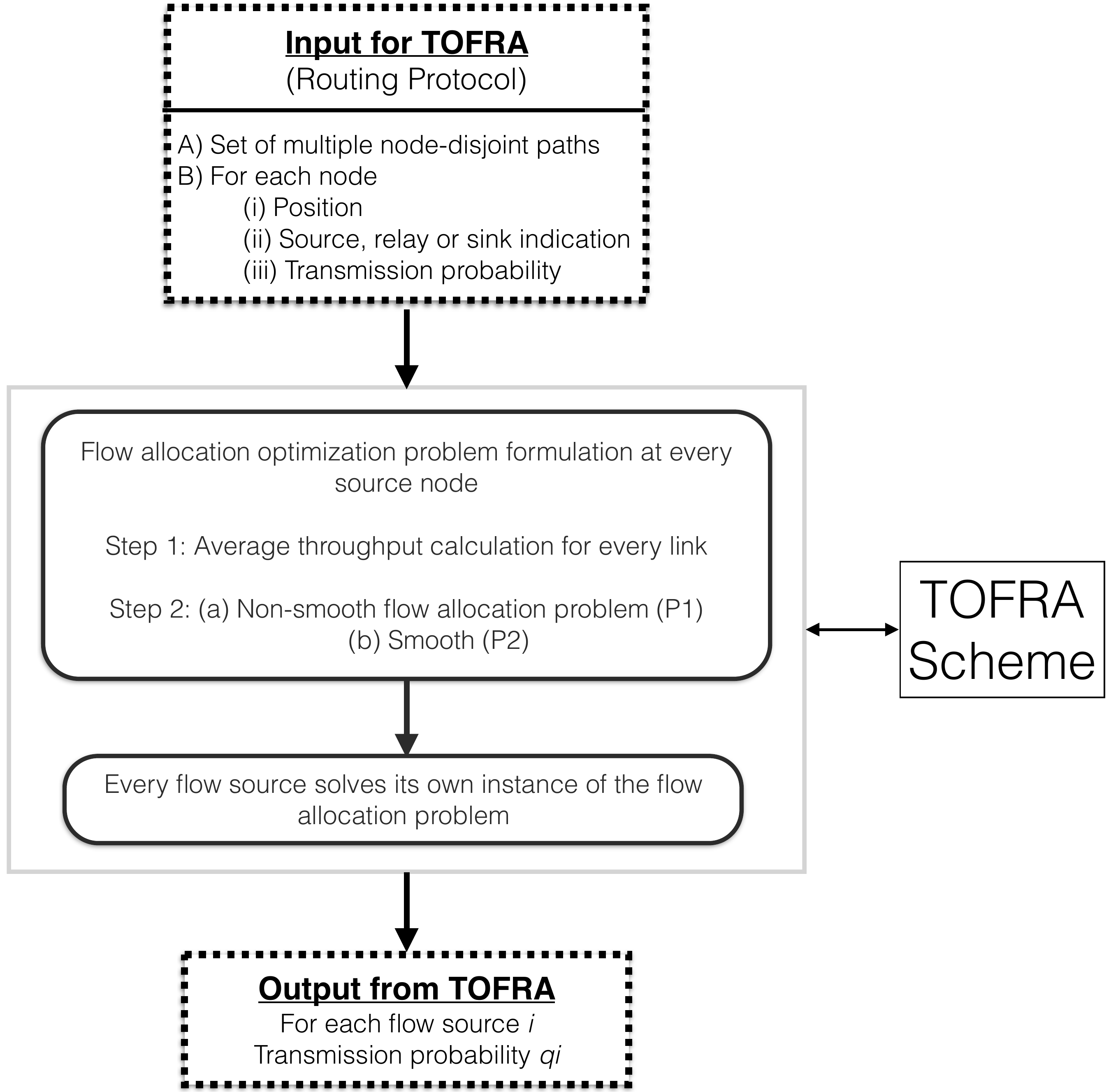}
\caption{Diagram of the proposed distributed flow allocation (TOFRA) scheme.}
\label{fig_state_diagram}
\end{figure}

In this section, the main concepts of the proposed flow allocation scheme are presented. First, the input required by the proposed
flow allocation scheme, from the routing protocol, is presented. This input is used for calculating flow rates that maximize
average aggregate flow throughput (AAT), on a distributed manner at each source. In Section~\ref{sec:analysis},  the analysis employed
for formulating flow rate allocation as an optimization problem is presented. The proposed scheme is demonstrated through a simple topology
in Section~\ref{sec:mpath_mhop_analysis}.

Introducing some of the notations, also required for the analysis presented in Section~\ref{sec:analysis}, we assume $m$ flows,
$f_{1}, f_{2},...,f_{m},$ that need to forward traffic to their destinations. For flow $f_{i}$, let $Src(f_{i})$ denote its source.
As also shown in Fig.~\ref{fig_state_diagram}, it is assumed that the routing protocol, provides each flow source, $Src(f_{i})$,
with a path to its corresponding destination, namely, $r_{i}$. It is further assumed that, these paths are node-disjoint thus, they do not share common nodes.
Implementing though, a routing protocol that identifies the path to be utilized, for each source and destination
pair, is out of the scope this study. However, it should be noted that, the proposed flow allocation scheme is independent of the routing
protocol implementation.

As already stated in the Introduction, the goal of the proposed scheme is to maximize AAT, while also providing bounded delay. Along this direction, the flow
rate with which, each source injects traffic, on the path employed, needs to be calculated. An additional requirement for the proposed scheme is that,
flow rate estimation should be distributed. For that reason, each flow source, should estimate independently of all other sources, the rate
of the flow injected, that contributes to achieving the maximum AAT for all flows. For doing so, each flow needs
to infer its own view of the network topology and derive its own instance of the \textit{flow allocation optimization problem},
presented in the next section. As Fig.~\ref{fig_state_diagram} shows, for deriving its own instance of the flow allocation problem,
each flow source is required to know for each other network node the following information: a) node position, b) type of node (source, relay, or sink),
and c) transmission probability for relays. Node positions for example, can be used to derive link distances, and thus success probability
for each link, based on (\ref{eq:succprob}) of Section~\ref{sec:system_model}. As also presented in Fig.~\ref{fig_state_diagram},
this information can be available for each node, throughout the network, through the routing protocol.
After having inferred its own instance of the flow allocation optimization problem, it can solve it and derive the flow rates
that flow sources should assign on each path, in order to collectively maximize AAT.

\subsection{Analysis}
\label{sec:analysis}

In this section, we present how aggregate throughput optimal flow rate allocation is formulated, at each flow source,
as an optimization problem, for random topologies. The suggested scheme is also demonstrated through a simple topology in order
to provide insights that are difficult to obtain through larger topologies.

\begin{table}[t]
\begin{center}
\begin{tabular}{|c|l|}
\hline
Notation & Definition\\ \hline
$V$ & Set of nodes. $|V|=N$\\ \hline
$f_{1}, f_{2},...,f_{m}$ & $m$ flows\\ \hline
$r_{i}$ & Path $i$ employed by flow $f_{i}$\\ \hline
$R = \lbrace r_{1}, r_{2}, ...,r_{m} \rbrace$ & Set of node disjoint paths \\ \hline
$|r_{i}|$ & Num of links in path $r_{i}$ \\ \hline
$I_{i,j}$ & Interfering nodes for link ($i$,$j$) \\ \hline
$I_{i,j}[n]$ & Id of n$^{th}$ interfering node for \\
&link (i,j) \\ \hline
$L_{i,j}=|I_{i,j}|$ & Number of nodes that interfere \\
& with transmissions on (i,j) \\ \hline
$Src(r_{k})$ & Source node of the $k^{th}$ flow \\ \hline
$P_{ r_{k} } = \prod_{ (i,j) \in r_{k} } p_{i/i}^{j}$ & End-to-end success probability \\
&for path $r_{k}$\\ \hline
$\bar{T}_{i,j}$ & Average throughput for (i,j) \\
&(Pkts/slot)\\  \hline
$\bar{T}_{r_{k}}$ & Average throughput for $k^{th}$ \\
&flow (Pkts/slot)\\ \hline
\end{tabular}
\end{center}
\caption{Flow allocation problem related notations}
\label{tab:notations}
\end{table}

The suggested method for formulating aggregate throughput optimal flow rate allocation as an optimization problem for random topologies is a
procedure consisting of two steps (also depicted in Fig.~\ref{fig_state_diagram}). We demonstrate this procedure assuming multiple
flows that are forwarded to the same destination.
The same analysis however can be applied for the case where multiple flows have different destination nodes.
First, the notations used in the analysis, are presented and are also summarized in Table~\ref{tab:notations}.
\textit{V} denotes the set of the nodes and $|V|=N$.
As also stated in the previous section, we assume $m$ flows $f_{1}, f_{2},...,f_{m},$ that need to forward traffic to the destination node $D$.
$R = \lbrace r_{1}, r_{2}, ...,r_{m} \rbrace$ represents the set of $m$ disjoint paths employed by these flows.
$|r_{i}|$ is used to denote the number of links in path $r_{i}$.
$I_{i,j}$ is the set of nodes that cause interference to packets sent from \textit{i} to \textit{j}.
For example, if all network nodes are assumed to contribute with interference to link $(i,j)$
and $j \neq D$, then $I_{i,j} = V \setminus \lbrace i, j, D \rbrace$ and thus, the set of nodes that cause interference to that link
has size $L_{i,j}=|I_{i,j}| = |V|-3$.
Further on, $Src(r_{k})$ is used to denote the source node of the $k^{th}$ flow, employing path $r_{k}$.
$\bar{T}_{i,j}$ and $\bar{T}_{r_{k}}$, denote the average throughput, measured in packets per slot, achieved by link $(i,j)$ and flow $f_{k}$
forwarded over path $r_{k}$, respectively.
Finally, $I_{i,j}[n]$ denotes the id of the n$^{th}$ interfering node for link $(i,j)$.

The first step of the suggested method, presented as Step $1$ in Fig.~\ref{fig_state_diagram},
consists of deriving the expression for the average throughput of a random link $(i,j)$.
\begin{equation}
\label{eq:process_step_1}
\bar{T}_{i,j}  =  \sum_{ l=0 }^{ 2^{ L_{i,j} }-1 } P_{i,j,l}  q_{i,j}  \prod_{n=1}^{ L_{i,j} } q_{I_{i,j}[n]}^{b(l,n)}  (1 - q_{I_{i,j}[n]} )^{1-b(l,n)},
\end{equation}
where
\begin{equation*}
\begin{aligned}
& q_{i,j}=\left\{\begin{matrix}
& q_{i} \quad & j =  D\\
& q_{i}(1-q_{j}) \quad & j\neq D\\
\end{matrix}\right.,\\
\end{aligned}
\end{equation*}
\begin{equation*}
\begin{aligned}
& P_{i,j,l}=p_{i/i\cup \{ I_{i,j}[n], \; \forall \; n: \; b(l,n) \neq0) \} }^{j}, \\
& b(l,n) = l \; \& \; 2^{n-1}, \; \text{\& is the logical bitwise AND operator.} \\
\end{aligned}
\end{equation*}

Average throughput for that link, $\bar{T}_{i,j}$, can be expressed
as the probability of having a successful packet reception over link $(i,j)$ and is given through (\ref{eq:process_step_1}).

The probability of a successfull packet transmission along link $(i,j)$ during a slot, depends among others, on the amount of received
interference. However, the amount of received interference also depends on the set of neighboring nodes that are transmitting
in each slot. Thus, the expression for a link's $(i,j)$ average throughput, requires the enumeration of all possible subsets of
interfering nodes. More precisely, assuming that $L(i,j)$ denotes the set of all nodes that cause interference to the transmissions
over the link $(i,j)$, all possible different subsets of active interfering nodes are $2^{L(i,j)}$. In each timeslot thus,
there are $2^{L(i,j)}$ different cases where a packet transmission may be successful for link $(i,j)$. The
probability of a successful packet transmission for link $(i,j)$, for all $2^{L(i,j)}$ cases, is captured through the
summation term in equation (\ref{eq:process_step_1}). In this equation, index $l$ runs from $0$ to $2^{L(i,j)}-1$, enumerating
all possible different subsets of active interfering nodes. Let $I_{l}^{i,j}$ for example, denote the $l^{th}$ subset of interfering
nodes for link $(i,j)$. The probability of a successful packet transmission over that link when all nodes in $I_{l}^{i,j}$ are
actvive at the same timeslot, is derived by considering the transmission probability of the nodes participating in $I_{l}^{i,j}$ and
the success probability of link $(i,j)$ given the interference by every transmitter in $I_{l}^{i,j}$. Note that, as also shown
in Table~\ref{tab_sys_mod_notations}, a node \textit{i}, is active during a slot with probability $q_{i}$. For flow originators,
$q_{i}$ denotes flow rate. As also described in Section \ref{sec:system_model}, transmission probability and position for every node,
can be periodically propagated to all other nodes, through routing protocol's control messages.
The success probability of link $(i,j)$, given the interference from nodes
in $I_{l}^{i,j}$, is captured through term $P_{i,j,l}$ in equation (\ref{eq:process_step_1}).
$P_{i,j,l}$ is in essence calculated through $p_{i/T}^{j}$ in equation \ref{eq:succprob} where $T$ in this case is the $I_{l}^{i,j}$.
Finally, it should be noted that, $b(l,n)$ in equation (\ref{eq:process_step_1}) becomes one if the $n^{th}$ node in $I_{i,j}$ is assumed
active in the $l^{th}$ subset examined.

For large networks though, enumerating all subsets of active transmitters may be computationally intractable.
In Section \ref{sec:evaluation} though, we explore a variant of the suggested flow allocation scheme, where only the \textit{k}
dominant interfering nodes are taken into account for expressing the throughput of link $(i,j)$.
As also discussed in that section, dominant interfering nodes for that link, are considered those that impose the most
significant amount of interference to packets received by \textit{j}.

The average aggregate throughput, achieved by all flows, is expressed through $\bar{T}_{aggr} = \sum_{ k= 1 }^{m} \bar{T}_{r_{k} }$,
where $\bar{T}_{r_{k}} = \underset{ (i,j) \in r_{k} }{ min }  \bar{T}_{i,j}$.
The second step of the suggested method, also depicted in Fig.~\ref{fig_state_diagram}, consists of maximizing the average aggregate throughput,
while also guaranteeing bounded packet delay which results in non-smooth optimization problem P1:
\begin{equation*}
\begin{aligned}
& \underset{S}{\text{Maximize}} \sum_{k=1}^{m} \underset{ (i,j) \in r_{k} }{ min } \bar{T}_{i,j} \quad \quad \quad \quad \quad \quad (P1)\\
& \text{s.t:} \\
& (S1):\;  0 \leq q_{Src(r_{k})} \leq 1, \; k=1,...,m\\
& (S2):\; \bar{T}_{ Src(r_{k}),i} \leq \bar{T}_{j,l}, \\
& \quad \quad \quad \lbrace \forall i,j,k,l :  (Src(r_{k}),i), \; (j,l) \in r_{k}, |r_{k}| > 1\\
& \quad \quad \quad k=1,...,m\rbrace, \\
\end{aligned}
\end{equation*}
where $S=\lbrace q_{Src(r_{k})}, \; k=1,...,m \rbrace$.
Constraint set S1 ensures that, the maximum data rate for any flow, does not exceed one packet per slot, while also allowing paths to remain unutilized.
Constraint S2 ensures that, the flow injected on each path, that is the throughput of that path's first link, is limited by the flow that
can be serviced by any subsequent link of that path. In this way, data packets are prevented from accumulating at the relay nodes,
guaranteeing thus, bounded packet delay.
For the rest of the paper, this constraint will be referred to as \textit{bounded delay constraint}.

P1 can be transformed to the following smooth optimization problem:
\begin{equation*}
\begin{aligned}
& \underset{S'}{\text{Maximize}} \sum_{k=1}^{m}\left\{\begin{matrix}
\bar{T}_{Src(r_{k}),D}, & |r_{k}|=1 \quad \quad \quad \quad \quad (P2)\\
q'_{Src(r_{k})}, & |r_{k}|>1 \quad \quad \quad \quad \quad \quad \quad \quad
\end{matrix}\right.\\
& s.t.: \\
& \quad \quad (S1): \; 0 \leq q_{Src(r_{k})} \leq 1, \; k=1,...,m\\
& \quad \quad(S2): \; \bar{T}_{ Src(r_{k}),i} \leq \bar{T}_{j,l}, \\
& \quad \quad \quad \quad \quad \lbrace \forall i,j,k,l :  (Src(r_{k}),i), \; (j,l) \in r_{k}, |r_{k}| > 1 \\
& \quad \quad \quad \quad \quad k=1,...,m \rbrace \\
& \quad \quad(S3): \; 0 \leq  q'_{Src(r_{k}) } \leq 1, \; \lbrace \forall k :  |r_{k}|>1 \rbrace \\
& \quad \quad(S4): \; q'_{Src(r_{k}) } \leq \bar{T}_{i,j}, \; \lbrace \forall i,j,k :  |r_{k}|>1, \; (i,j) \in r_{k} \rbrace, \\
\end{aligned} \\
\end{equation*}
where $S'=\lbrace q_{Src (r_{k} )}, \;  k=1,...,m \rbrace \cup \lbrace q'_{src (r_{k} )} : \;  |r_{k}|>1 \rbrace$.
For the rest of the paper, we will refer to optimization problem P2 above as, the \textit{flow allocation optimization problem}.

\subsection{Throughput optimal flow rate allocation: An illustrative scenario}
\label{sec:mpath_mhop_analysis}

\begin{figure}
\centering
\includegraphics[scale=0.4]{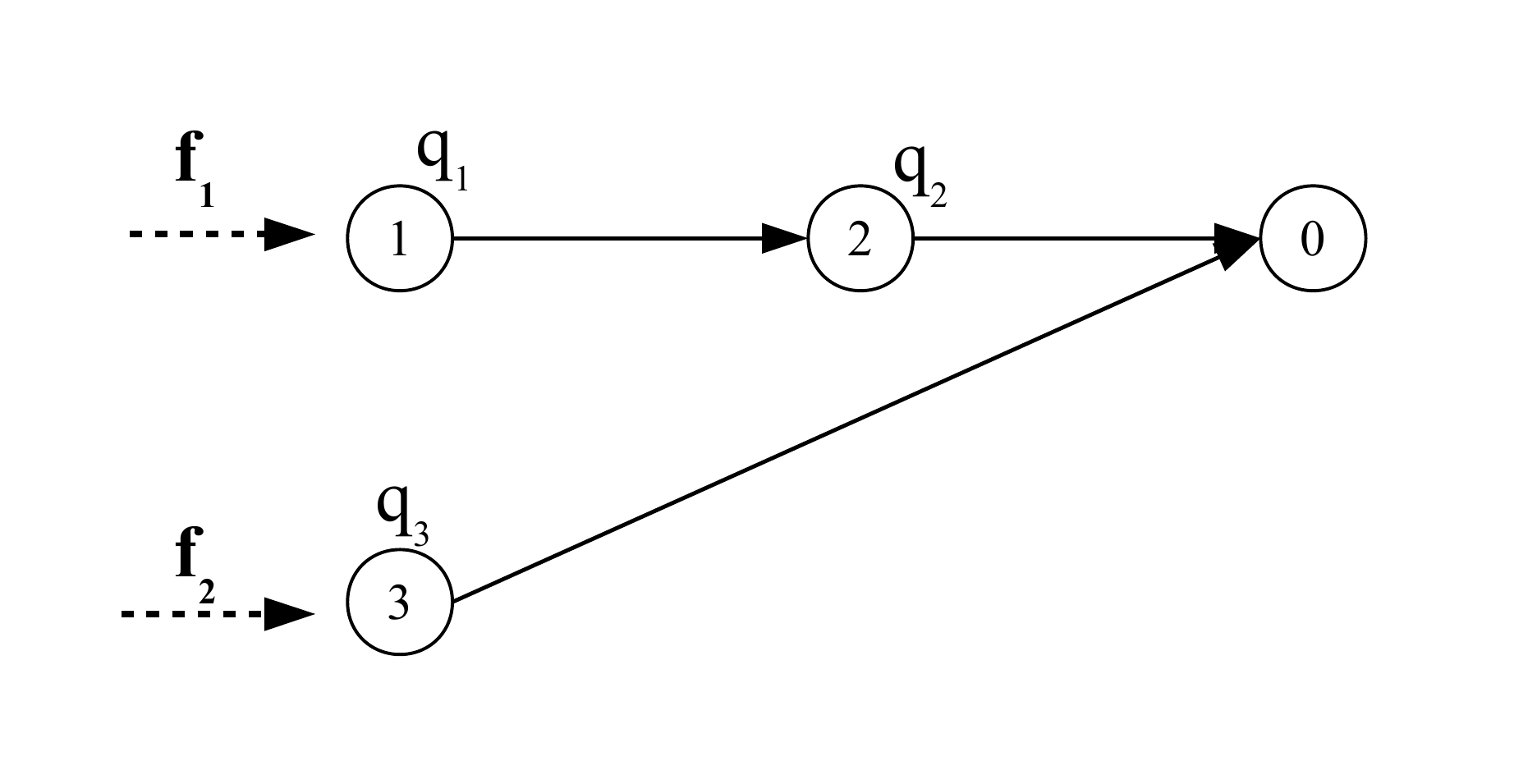}
\caption{An illustrative topology.}
\label{fig:mpath_mhop}
\end{figure}

We consider the simple topology presented in Fig. \ref{fig:mpath_mhop}.
This simple topology is used to illustrate key features of the proposed scheme and also to provide insights concerning
the relation among the flow allocation on each path, interference, and SINR threshold employed.
Such insights are hard to obtain from more complex scenarios.
Two flows namely, $f_{1}$ and $f_{2}$, originating from nodes $1$, and $3$, are forwarded to destination node $0$
through paths $r_{1}$: $1 \rightarrow 2 \rightarrow 0$ and $r_{2}$: $3 \rightarrow 0$, respectively.
We further assume that, transmissions on a specific link, cause interference to all other links.
Before presenting each link's average throughput, consider link $(2,0)$ as an example.
Transmitters that cause interference to packets sent from $2$ to $0$, constitute set $I_{2,0} = \lbrace 1,3 \rbrace$ and thus $L_{2,0}=2$.
There are three possible subsets of nodes that may cause interference on link $(2,0)$ : $\lbrace 1  \rbrace , \lbrace 3  \rbrace ,\lbrace 1,3  \rbrace $.
When $l=3$, in (\ref{eq:process_step_1}), it indicates the third subset of interfering nodes, with $b(l,n)$ becoming one, for both $n=1$ and $n=2$.

The average throughput per link is presented in (\ref{eq:multi_path_one_relay_linkthr}a)-(\ref{eq:multi_path_one_relay_linkthr}c).
\begin{subequations}
\label{eq:multi_path_one_relay_linkthr}
\begin{align}
{\bar{T}}_{1,2} &= q_{1}(1-q_{2})(1-q_{3})p_{1/1}^{2} + q_{1}(1-q_{2})q_{3}p_{1/1,3}^{2} \\
{\bar{T}}_{2,0} &= q_{2} (1-q_{1}) (1-q_{3}) p_{2/2}^{0} + q_{2} q_{1} (1-q_{3}) p_{2/2,1}^{0} \nonumber \\ &+ q_{2} (1-q_{1}) q_{3} p_{2/2,3}^{0} + q_{2} q_{1} q_{3} p_{2/1,2,3}^{0}\\
{\bar{T}}_{3,0} &= q_{3} (1-q_{1}) (1-q_{2}) p_{3/3}^{0} + q_{3} q_{1} (1-q_{2}) p_{3/1,3}^{0} \nonumber \\ &+ q_{3} (1-q_{1}) q_{2} p_{3/2,3}^{0} + q_{3} q_{1} q_{2}  p_{3/1,2,3}^{0}
\end{align}
\end{subequations}

Recall that, $q_{1}$ and $q_{3}$, denote the data rates for flows, $f_{1}$ and $f_{2}$, respectively.
Aggregate average throughput achieved by all flows can be expressed through (\ref{eq:one_relay_aggr_throughput}).
\begin{equation}
\label{eq:one_relay_aggr_throughput}
\begin{aligned}
& {\bar{T}}_{aggr} ={\bar{T}}_{r_{1}}+{\bar{T}}_{r_{2}}, \quad where, \\
& {\bar{T}}_{r_{1}} = min\{ {\bar{T}}_{1,2},{\bar{T}}_{2,0} \}, \quad {\bar{T}}_{r_{2}} = {\bar{T}}_{3,0}
\end{aligned}
\end{equation}

Average aggregate throughput-optimal flow rate allocation, consists of identifying rates, $q_{1}$ and $q_{3}$, that
maximize average aggregate throughput, while also guaranteeing bounded packet delay.
These rates can be found by solving the following optimization problem:
\begin{equation*}
\begin{aligned}
& \underset{q_{1},q_{3}}{\text{Maximize}}
& & {\bar{T}}_{30}+min\{ {\bar{T}}_{12},{\bar{T}}_{20} \}\\
& \text{subject to}
& & 0 \leq q_{i} \leq 1, \; i \in \lbrace 1,3 \rbrace \quad (g1)-(g4)\quad \quad \quad \\
& & & {\bar{T}}_{12}\leq {\bar{T}}_{20} \quad \quad \quad \quad\quad \quad(g5)\\
\end{aligned}
\end{equation*}

Constraint (g5) constitutes the bounded delay constraint for path $r_{1}$.
The above non-smooth optimization problem can be transformed to the following smooth optimization problem:
\begin{equation*}
\begin{aligned}
& \underset{q_{1}^{'},q_{1},q_{3} }{\text{Maximize}}
& & {\bar{T}}_{30}+q_{1}^{'} \\
& \text{subject to}
& & 0 \leq q_{i}\leq 1, \; i \in \lbrace 1,3 \rbrace \quad & (g1) &-(g4)\\
& & & {\bar{T}}_{12}\leq {\bar{T}}_{20}, & (g5)\\
& & & q_{1}^{'} \leq {\bar{T}}_{12}, & (g6) & \quad \quad \quad \quad (P3)\\
& & & q_{1}^{'} \leq {\bar{T}}_{20}, & (g7)\\
& & & 0 \leq q_{1}^{'} \leq 1 & (g8)&-(g9)\\
\end{aligned}
\end{equation*}

Transforming the above optimization problem in the standard form, the function over $q_{1}^{'}, q_{1},q_{3}$ related to constraint (g5),
$g_{5}(q_{1}^{'}, q_{1},q_{3})={\bar{T}}_{12} - {\bar{T}}_{20}$ is non-convex and thus the problem is non-convex.

Before presenting simulation results for random wireless scenarios, we further motivate flow rate allocation on multiple paths, using
numerical results derived from the simple topology depicted in Fig~\ref{fig:mpath_mhop}. Let $d(i,j)$ denote the distance between nodes $i$ and $j$.
Let also, $P_{ r_{k} } = \prod_{ (i,j) \in r_{k} } p_{i/i}^{j}$, denote the \textit{end-to-end success probability}, for path $r_{k}$.
For the illustrative purpose of this section, we assume that $d(1,2)=d(2,0)=d(3,1)=d$, $d(3,0)=\sqrt 5d$, $d(3,2) = \sqrt 2 d$, where $d=400m$.
Further on, the path loss exponent assumed is $3$, while the transmission probability for relay node $2$, is $0.5$.
Flow rates, $q_{1}$ and $q_{3}$, that achieve maximum average aggregate throughput (AAT), for SINR threshold values $\gamma = \lbrace 0.25, 0.5,...,2 \rbrace$,
are estimated by solving the optimization problem (P3) using the simulated annealing technique.
It should be noted that, multi-hop path $r_{1}$: $1 \rightarrow 2 \rightarrow 0$ exhibits higher end-to-end success probability
than path $r_{2}$: $3 \rightarrow 0$, for all $\gamma$ values considered.
\begin{figure}
\centering
\includegraphics[scale=0.3]{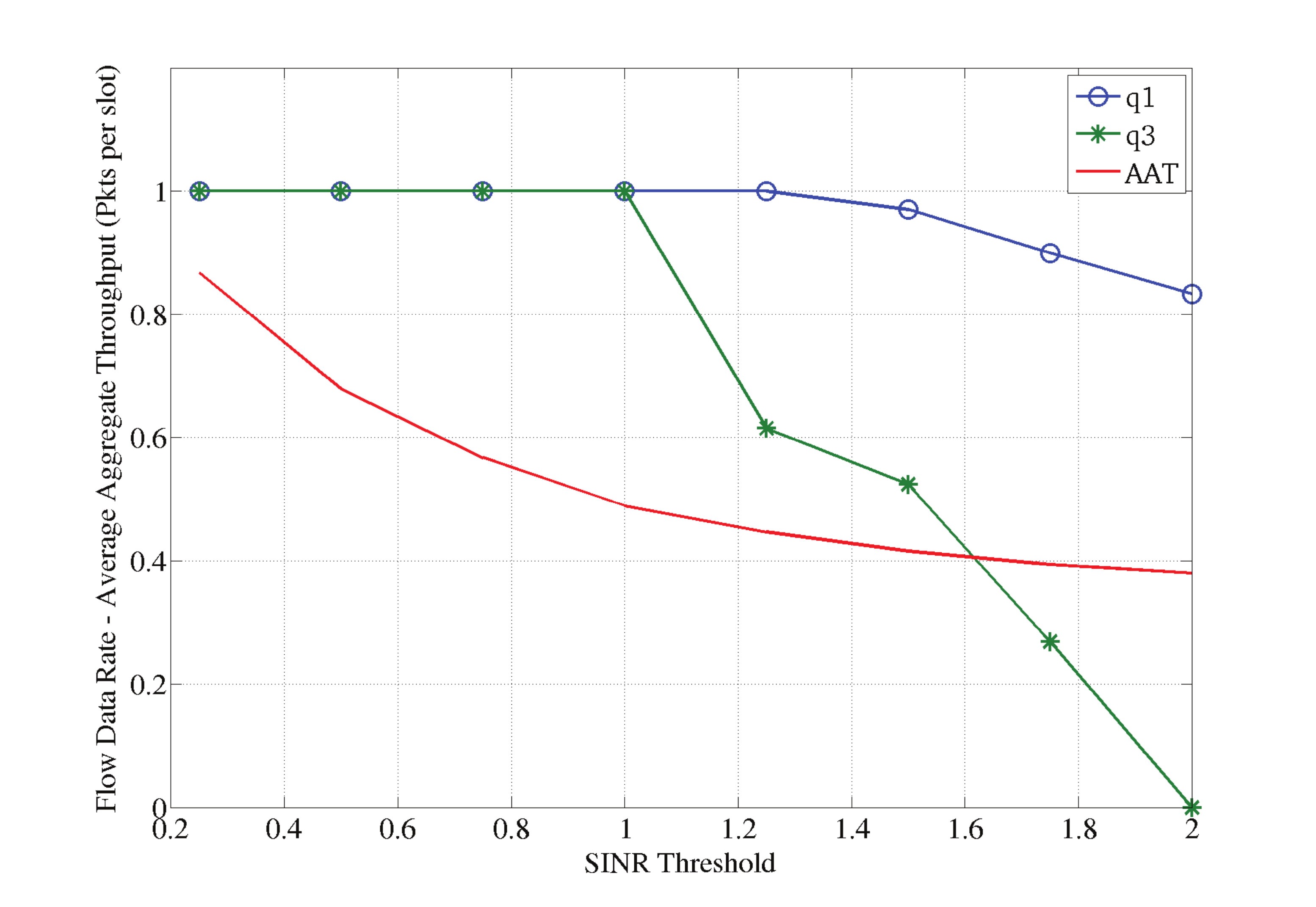}
\caption{Optimal Flow rates and Average Aggregate Throughput achieved.}
\label{fig:num_res_toy_topo}
\end{figure}

In Fig. \ref{fig:num_res_toy_topo}, we present throughput optimal flow rates assigned on paths, $r_{1}$ and $r_{2}$, along with the average
aggregate throughput achieved (AAT), for the aforementioned $\gamma$ values.
As this figure shows, the maximum AAT is achieved by full rate utilization of both paths, for SINR threshold values up to $1.0$, suggesting
that inter-flow interference is balanced by the gain in throughput.
For SINR threshold values larger than $1.0$, utilization of path $r_{2}$, which exhibits lower performance in terms of end-to-end success probability, declines.
This is due to the fact that, for large SINR threshold values, the effect of interference imposed on path $r_{1}$ becomes more significant.
At the same time, the flow forwarded through path $r_{2}$, manages to deliver only a small portion of its traffic to destination node $0$.

\section{Evaluation}
\label{sec:evaluation}

\subsection{Simulation setup}
\label{sec:sim_setup}

The proposed throughput optimal flow rate allocation (TOFRA) scheme, is evaluated using the network simulator Ns2, version 2.34 \cite{ref:ns2}.

\begin{table}[t]
\begin{center}
\begin{tabular}{ll}
\hline
Parameter & Value\\ \hline
Max Retransmit Threshold & $3$\\
Transmit Power & $0.1$ Watt\\
Noise power & $7\times10^{-11}$ Watt\\
Contention Window & 5\\
Packet size & $1500$ Bytes\\
Path loss exponent & 4 \\
\hline
\end{tabular}
\caption{Network parameters used for deriving numerical and simulation results}
\label{tab:param_simul}
\end{center}
\end{table}

Concerning medium access control, a slotted aloha-based MAC layer is implemented.
Transmission of data, routing protocol control and ARP packets is performed at the beginning of each slot, without performing carrier sensing prior to transmitting.
Acknowledgements for data packets are sent immediately after successful packet reception, while failed packets are retransmitted.
Slot length, $T_{slot}$, is expressed through: $T_{slot} = T_{data} + T_{ack} + 2D_{prop}$, where $T_{data}$ and $T_{ack}$, denote
the transmission times for data packets and acknowledgements (ACKs), while $D_{prop}$ denotes the propagation delay.
It should be noted that, all packets have the same size, shown in Table~\ref{tab:param_simul}.
All network nodes, apart from sources of traffic, select a random number of slots before transmitting, drawn
uniformly from $[0,CW]$. The contention window (CW) is fixed for the whole duration of the simulation and equal to $5$.

As far as physical layer is concerned, all data packets are successfully decoded if their received SINR exceeds the SINR threshold.
The received SINR for each packet is calculated through (\ref{eq:sinr_thres}). The path loss exponent is assumed to be $\alpha=4$.
Transmitters during each slot, that are considered to cause interference, are those transmitting data packets, or routing protocol control packets.
All nodes use the same SINR threshold, transmission rate, and channel. Transmission power and noise is $0.1$ Watt and $7 \times 10^{-11}$ Watt, respectively.

As far as routing is concerned,  a multipath, source-routed link-sate routing protocol based on UM-OLSR \cite{ref:um-olsr}, is implemented.
Hello and Topology Control (TC) messages are periodically propagated throughout the network.
Each topology control message may carry the following information: a) transmission probability b) position, and
c) an indication of whether it is a flow originator, relay or, sink.
As also discussed in Section~\ref{sec:system_model}, transmission probabilities are assumed to be fixed for relay nodes, since contention
window (CW) remains fixed for the whole simulation period. For flow originators, transmission probabilities are estimated
by solving the corresponding version the flow allocation optimization problem (presented in Section \ref{sec:analysis}), using the simulated annealing technique.
Identification of the multiple, node-disjoint paths to be utilized by the various flows, can be performed at flow sources each time a new TC message is received.
However, the main focus of the study, is not on identifying the set of paths that should be employed, but
on how should these paths be utilized in order to maximize average aggregate throughput for all flows, with \textit{how}
referring to the amount of flow assigned on each path.
We thus, employ a simple algorithm that provides traffic sources with multiple, link-disjoint, least-cost paths.
The multipath set is populated on an iterative manner. On each iteration, a specific flow's source and destination node are considered.
The graph inferred from TC messages, is searched for a least cost path, using the Dijkstra algorithm.
The nodes participating in the path identified, are removed and the search process continues with the next flow's source and destination node.
In this way, the multipath set consists of node disjoint paths.

As Fig.~\ref{fig_state_diagram} also shows, upon each TC message reception, and having inferred the node disjoint paths utilized, each source
may employ TOFRA to infer a topology-specific instance of the flow allocation optimization problem.
In this way, each flow source can independently of all other sources,
identify the flow rates that collectively achieve maximum AAT. That is, after all the required information has been propagated to flow sources,
then, each of them, can separately formulate and solve the corresponding flow allocation optimization problem.
According to this process, flow rates are estimated on a distributed manner, for all flow originators.

As far as queues at the relay nodes are concerned, two variants of the proposed TOFRA scheme are simulated. The first variant
follows the assumption of saturated queues in the analysis, while in the second variant, queues are not assumed to be saturated.
The goal of this process is to gradually evaluate, whether the suggested flow allocation scheme, accurately captures the average
aggregate throughput (AAT) observed in the simulation results. With the first variant, we explore whether
the model for the AAT employed by the suggested scheme, accurately captures the effect of random access and interference on AAT.
The second variant, explores the effect of the assumption concerning saturated queues on accurately capturing the AAT observed in the simulated scenarios.
In order to implement the first variant, the following patch is required
in Ns2: whenever a relay node \textit{i}, successfully receives a packet destined for a next hop \textit{j}, it buffers the full
header of the packet. Then, if the queue for the next hop gets empty during a subsequent slot, it creates a new dummy packet
with a dummy payload and adds the header buffered. Dummy packets are not taken into account for average aggregate flow throughput calculation.

\begin{figure}
\centering
\includegraphics[scale=0.2]{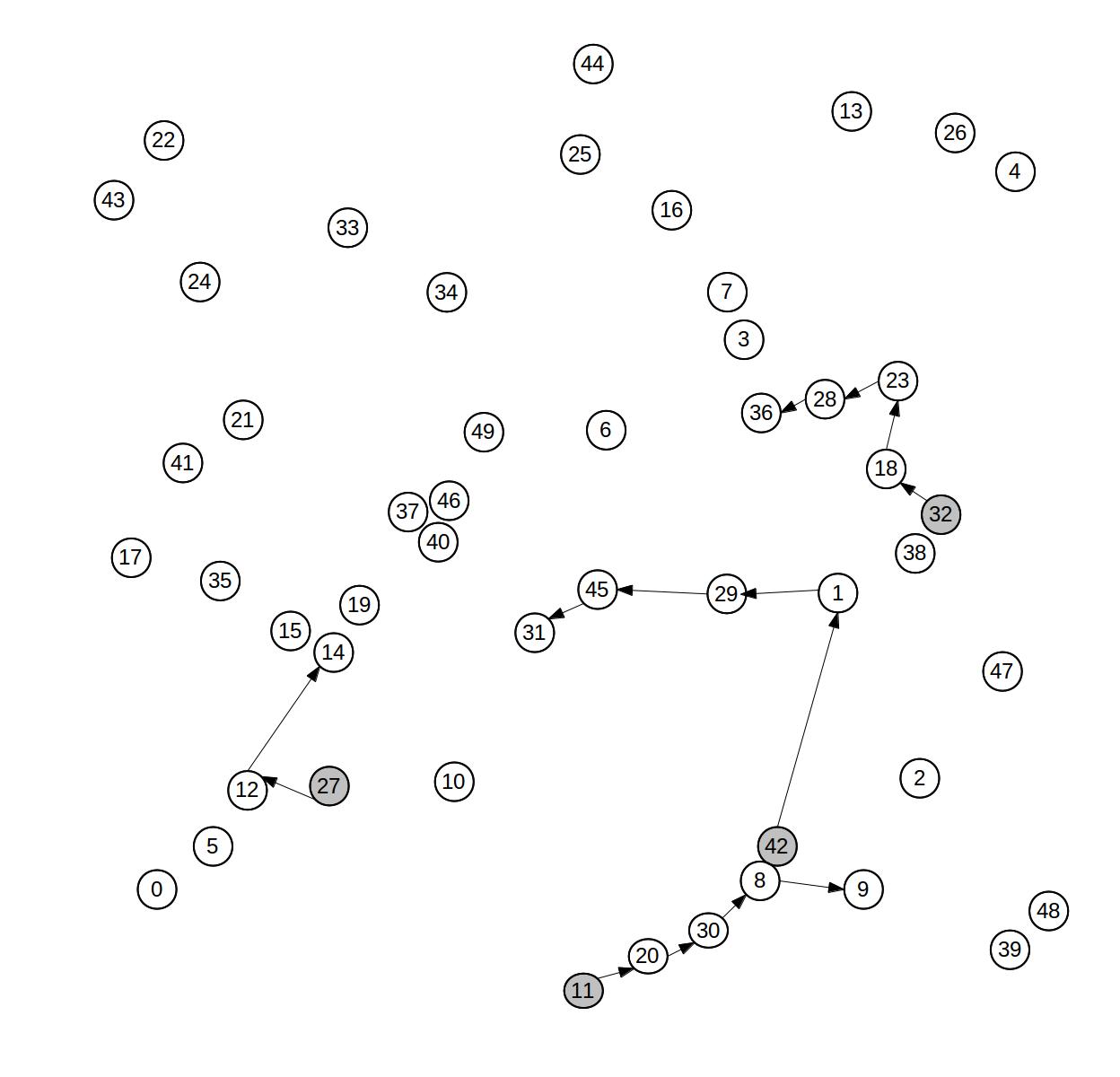}
\caption{Illustrative random wireless scenario.}
\label{fig:demo_rand_wireless_scen}
\end{figure}

For the purposes of the evaluation process, ten different wireless scenarios are generated.
For these scenarios, $50$ nodes are uniformly distributed, over an area of $500$m $\times$ $500$m.
The number of flows generated, along with the source and destination node for each flow, are selected
randomly. A maximum number of ten flows is allowed for each scenario and the simulation time is $20.000$ slots.
Traffic sources generate UDP unicast flows and are kept backlogged for the whole simulation period.
\begin{figure}
\centering
\includegraphics[scale=0.3]{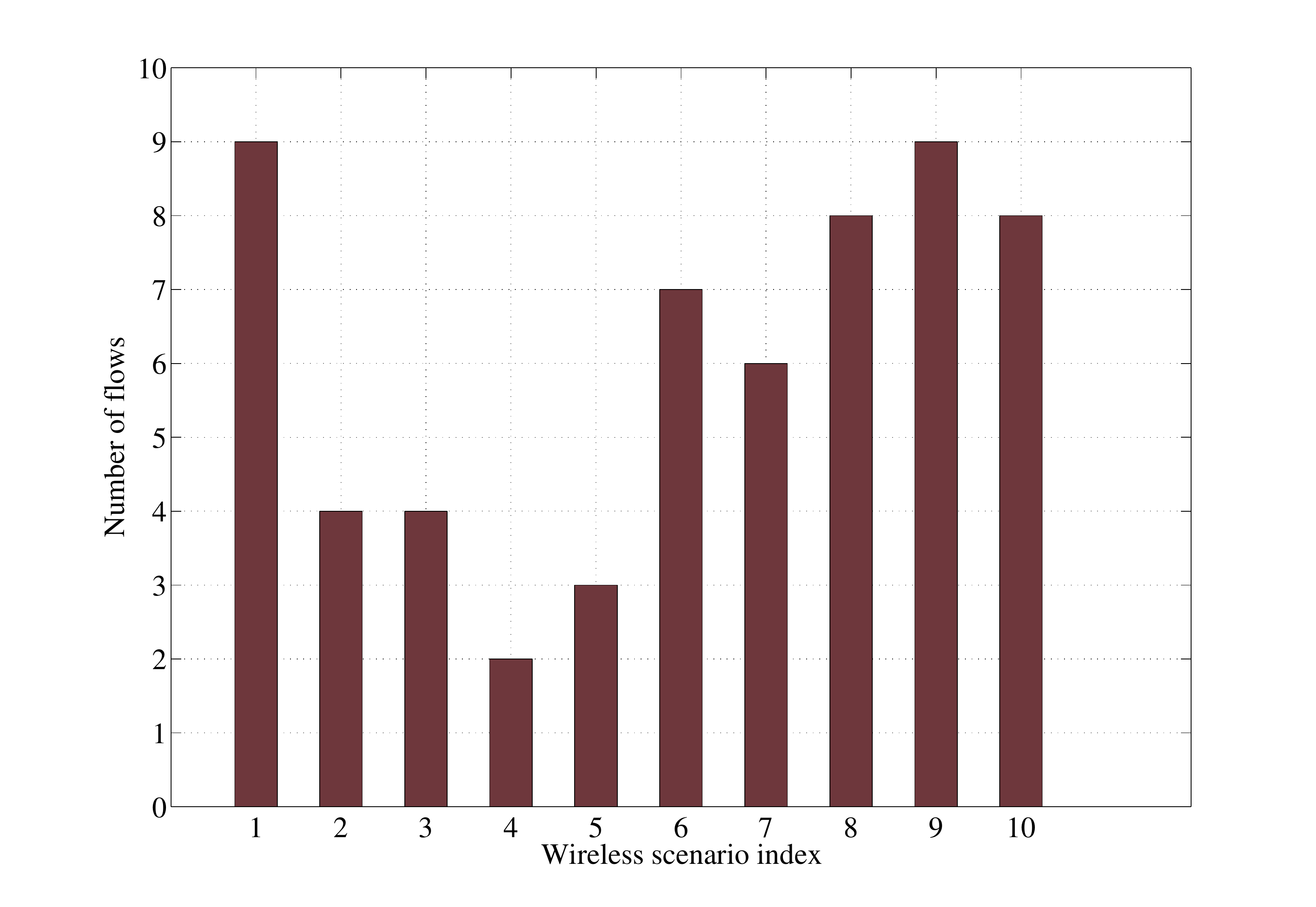}
\caption{Number of flows per wireless scenario.}
\label{fig:num_of_flows}
\end{figure}

Fig. \ref{fig:num_of_flows}, presents the number of flows generated, for each one of the ten wireless scenarios employed.
Fig. \ref{fig:demo_rand_wireless_scen}, depicts one such wireless scenario, including four flows.
The source and destination nodes for flow $f_{1}$, are $42$ and $31$, respectively. The corresponding source and destination
nodes for flows $f_{2}$, $f_{3}$, and $f_{4}$, are: $(11,9)$, $(32,36)$, and $(27,14)$.
Before employing the suggested flow allocation scheme, for determining the flow that should be assigned on each path,
multiple node-disjoint paths need to be identified, one for each flow.
As also shown in this figure, the paths employed for these flows are: $r_{1}: 42 \rightarrow 1 \rightarrow 29 \rightarrow 45 \rightarrow 31$,
$r_{2}: 11 \rightarrow 20 \rightarrow 30 \rightarrow 8 \rightarrow 9$, $r_{3}: 32 \rightarrow 18 \rightarrow 23 \rightarrow 28 \rightarrow 36$,
and $r_{4}: 27 \rightarrow 12 \rightarrow 14$.
Note that the output of the TOFRA scheme is flow rates $q_{42}$, $q_{11}$, $q_{32}$, and $q_{27}$, that will provide with the maximum average aggregate throughput,
while also guaranteeing bounded packet delay.
In order to capture the effect of interference on success probability and thus, on throughput, different $\gamma$ values are considered.
The corresponding values are $0.5, 1.0, 1.5$, and $2.0$, respectively.

\subsection{Simulation Results}
\label{sec:sim_results}

\begin{figure}
\centering
\includegraphics[scale=0.30]{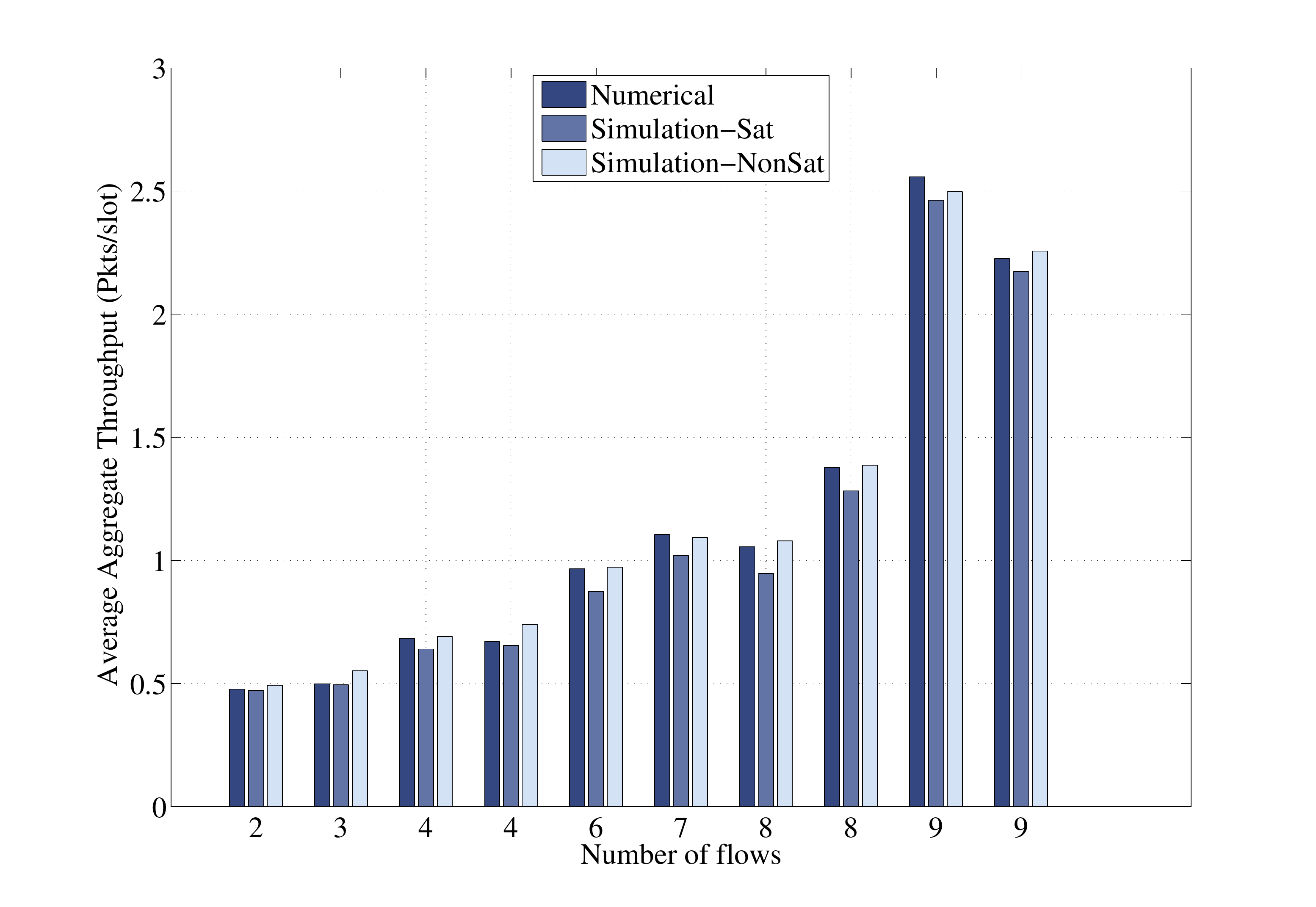}
\caption{Average Aggregate Throughput: Numerical vs. Simulation results for $\gamma$ $=$ $0.5$.}
\label{fig:eval_part1_g05}
\end{figure}
\begin{figure}
\centering
\includegraphics[scale=0.30]{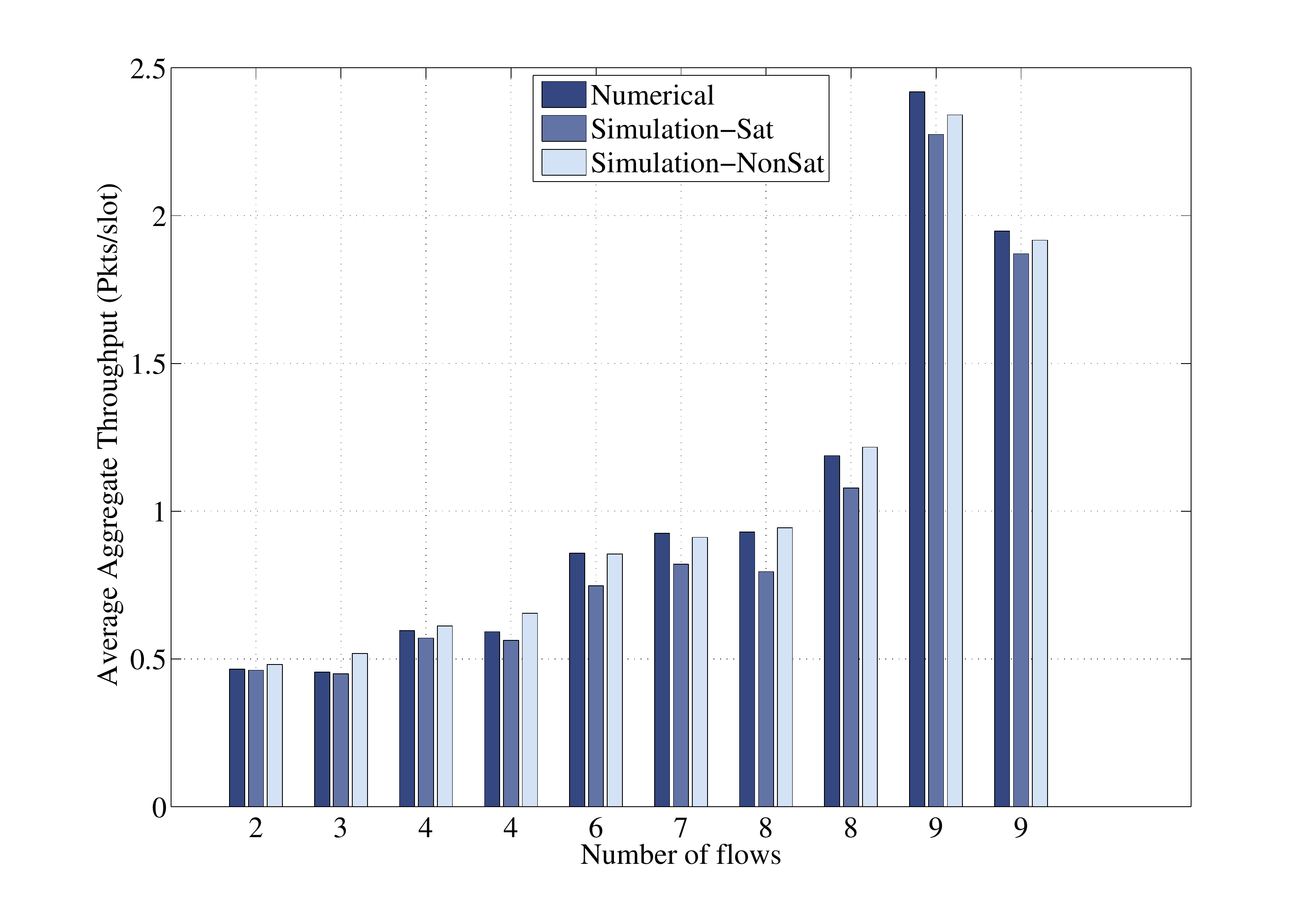}
\caption{Average Aggregate Throughput: Numerical vs. Simulation results for $\gamma$ $=$ $1.0$.}
\label{fig:eval_part1_g10}
\end{figure}
\begin{figure}
\centering
\includegraphics[scale=0.30]{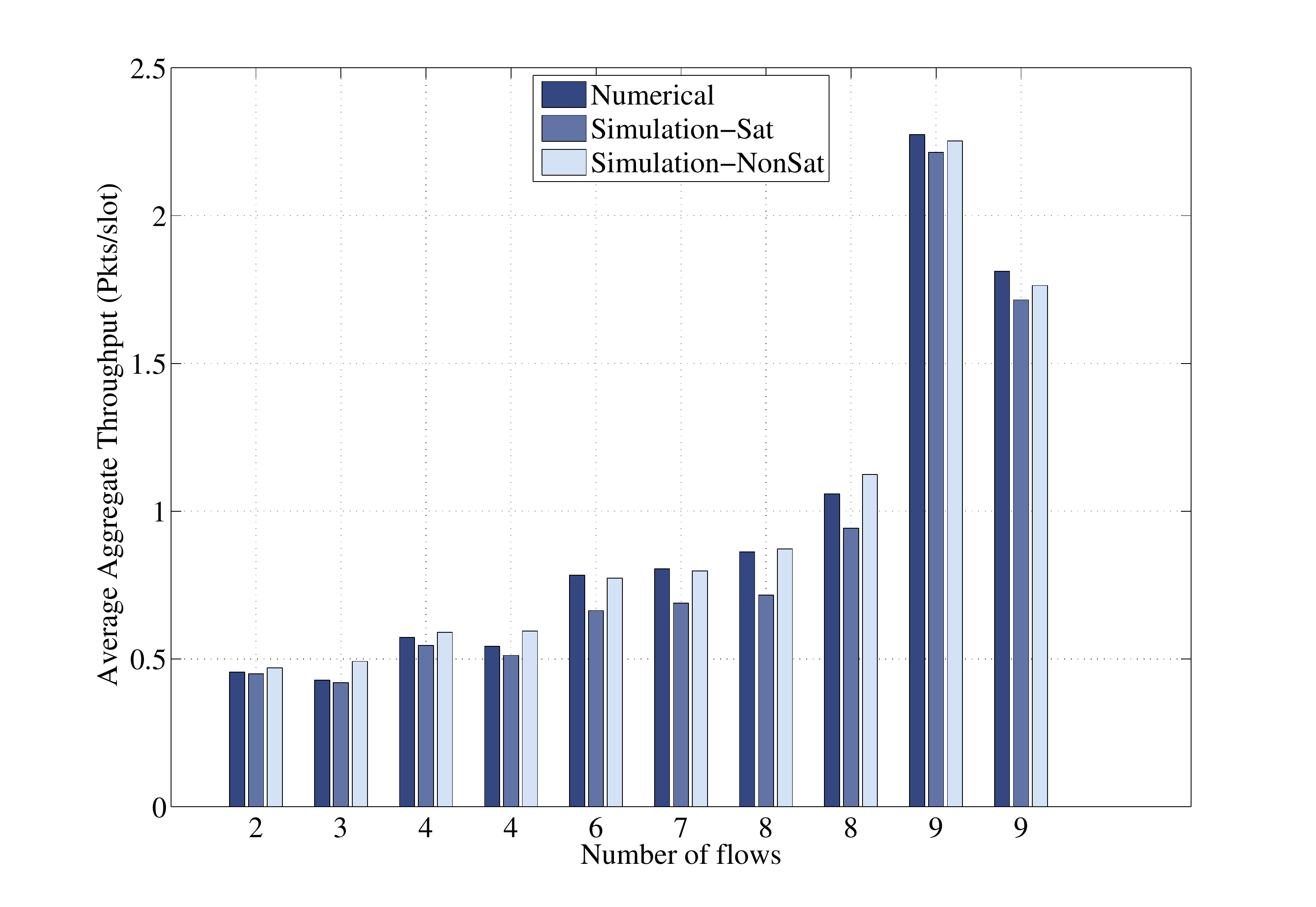}
\caption{Average Aggregate Throughput: Numerical vs. Simulation results for $\gamma$ $=$ $1.5$.}
\label{fig:eval_part1_g15}
\end{figure}
\begin{figure}
\centering
\includegraphics[scale=0.30]{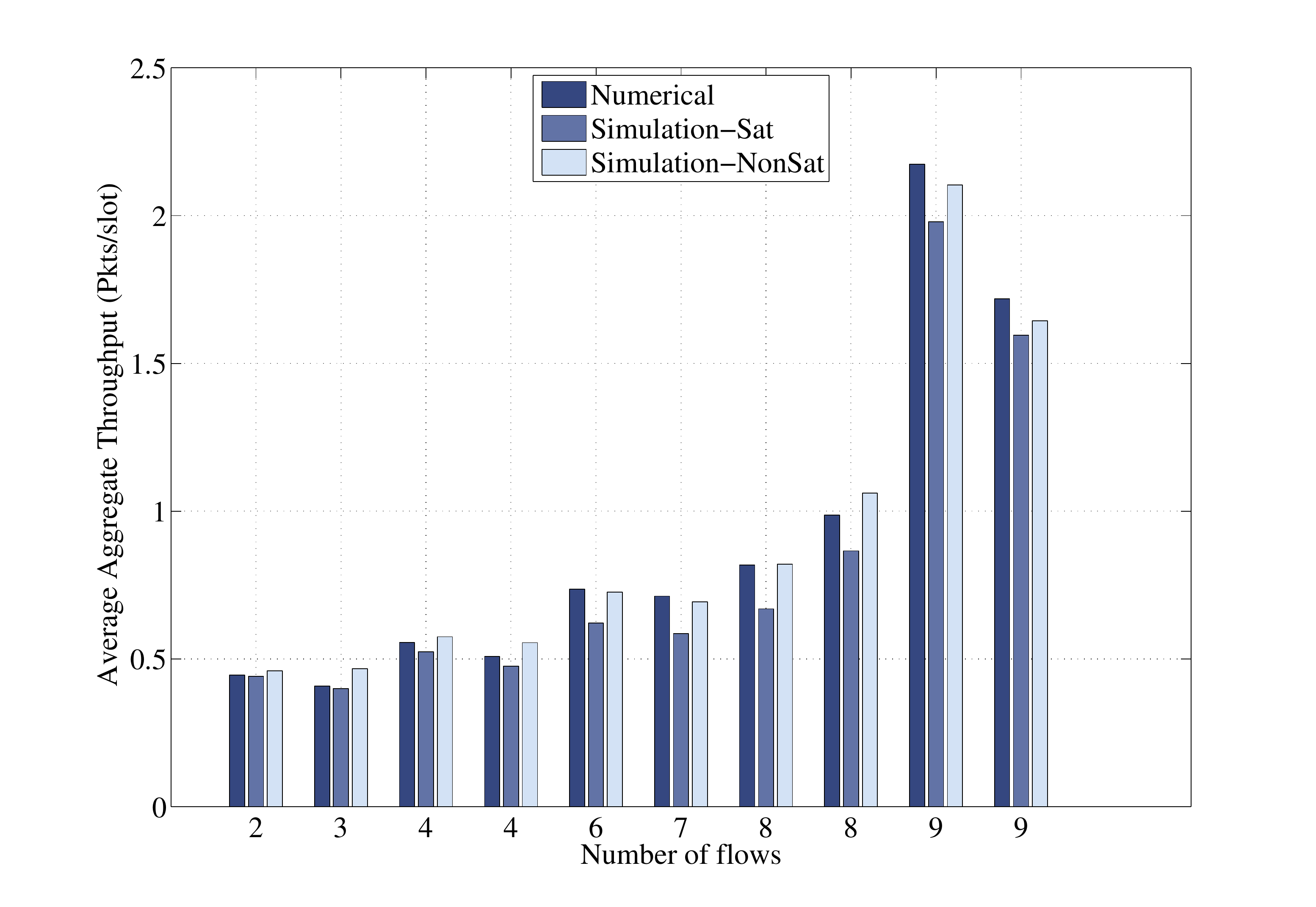}
\caption{Average Aggregate Throughput: Numerical vs. Simulation results for $\gamma$ $=$ $2.0$.}
\label{fig:eval_part1_g20}
\end{figure}

The evaluation process consists of three parts. In the first one, we explore whether the model employed by the proposed
flow allocation scheme (TOFRA), accurately captures the average aggregate throughput (AAT) observed in the simulated scenarios.
To introduce the notation used in the figures below, simulation results for the TOFRA variant that is simulated assuming saturated queues,
are labelled as \textit{Simulation-Sat}. Simulation results for the TOFRA variant where the assumption of
saturated queues is removed, are labelled as \textit{Simulation-NonSat}.

Figs.~\ref{fig:eval_part1_g05} to \ref{fig:eval_part1_g20}, compare numerical with simulation results, concerning AAT, for SINR threshold values
$0.5$, $1.0$, $1.5$, and $2.0$ and the ten different wireless scenarios explored.
Simulated results for both TOFRA variants are presented.
For the case of the TOFRA variant, where queues for relay nodes are kept backlogged for the whole simulation period, the average
deviation over all simulated scenarios, between the numerical and simulation results, is $5.5\%$, $7.6\%$, $9.0\%$, and $10.9\%$,
respectively, for the four SINR threshold values considered. In all the scenarios and for all the $\gamma$ values considered,
the model employed by the TOFRA scheme overestimates the AAT observed in the simulation results. There are two reasons for this overestimation.
The first one, is related to the maximum retransmit threshold.
In the analysis employed, its effect is disregarded and thus, no packet is dropped after exceeding a certain number of failed retransmissions.
In the simulated results however, it is set to $3.0$, which means that a packet that is unsuccessfully transmitted for three times, it will be
dropped. If there is no other packet available in the transmitter's queue, a dummy packet will be inserted (in case where TOFRA is simulated
with the saturated queues assumption) instead. Dummy packets however, are not taken into account for AAT calculation.
More on the effect of maximum retransmit threshold on TOFRA's AAT, consider scenario, $8$ with $\gamma=1.0$, as an example.
Simulated AAT for the proposed scheme, when queues are saturated and the maximum retransmit threshold is $3$, is $16.6\%$
lower than the corresponding numerical value. When the corresponding scenario is simulated with
an infinite value for the maximum retransmit threshold, the corresponding deviation between numerical and simulated AAT
drops to $1.9\%$.
The second reason, for the overestimation of the AAT observed in the simulated scenarios is the following:
in the analysis, it is assumed that whenever a packet is transmitted it is a packet carrying data.
In the simulated scenarios however, all nodes either perform periodic emission of routing protocol's control messages, or forward
received control packets.
This means that, specific slots are spent carrying routing protocol's control messages, instead of data packets, resulting
in our analysis overestimating the AAT observed in the simulated results. The second reason for AAT overestimation though, is less
important due to the large intervals over which control packets are generated and the small number of nodes participating
in the multipath set.

Figs.~\ref{fig:eval_part1_g05} to \ref{fig:eval_part1_g20}, also compare the AAT achieved by TOFRA model and the simulated one,
when the assumption of saturated queues at the relays is removed.
There are three reasons that shape the gap between the numerical
and the simulated AAT for TOFRA, when queues are not saturated, with all three reasons stemming from analysis' assumptions.
The first two reasons were described in the previous paragraph and result in our analysis overestimating the AAT observed in the simulated results.
The third reason has an opposite effect on AAT and is related to the saturated queues assumption present in the analysis.
According to this assumption, whenever a relay node attempts to transmit a packet there is always one available for transmission in its queue.
In the simulated scenarios however, this is not always the case.
As a result, the actual interference experienced by transmissions along
a link, is lower than the one assumed in the analysis and thus, the actual average throughput for a link may be higher than the one
calculated by the analysis applied. The effect of this is that, the model employed may underestimate the average throughput of a specific links
and thus, may underestimate the AAT. For each $\gamma$ value employed, the average deviation between numerical and simulated results,
concerning AAT, is estimated over all ten traffic scenarios explored.
The corresponding average deviation values are $3.1\%$, $3.7\%$, $4.0\%$, and $4.7\%$, for $\gamma=0.5, 1.0, 1.5, 2.0$, respectively.
Note that, for each wireless scenario, the absolute value of the deviation of simulated from numerical AAT is considered.
It is interesting to note that, the deviation between numerical and simulated results, concerning AAT, is lower
for the case where the assumption of saturated queues is removed. This is however, due to the contradictory effects on AAT, between
the assumption of saturated queues and the assumptions concerning the maximum retransmit threshold, and the occupation of certain slots
by routing protocol's control traffic.

For the rest of the evaluation process, only queues of flow originators will be kept backlogged for the whole simulation period. Queues for relay
nodes may be empty during a specific slot.
Finally, the minimum and maximum AAT variance value, over all traffic scenarios explored and the four $\gamma$ values considered, are
$10^{-4}$ and $10^{-3}$, respectively.

\begin{figure}
\centering
\includegraphics[scale=0.30]{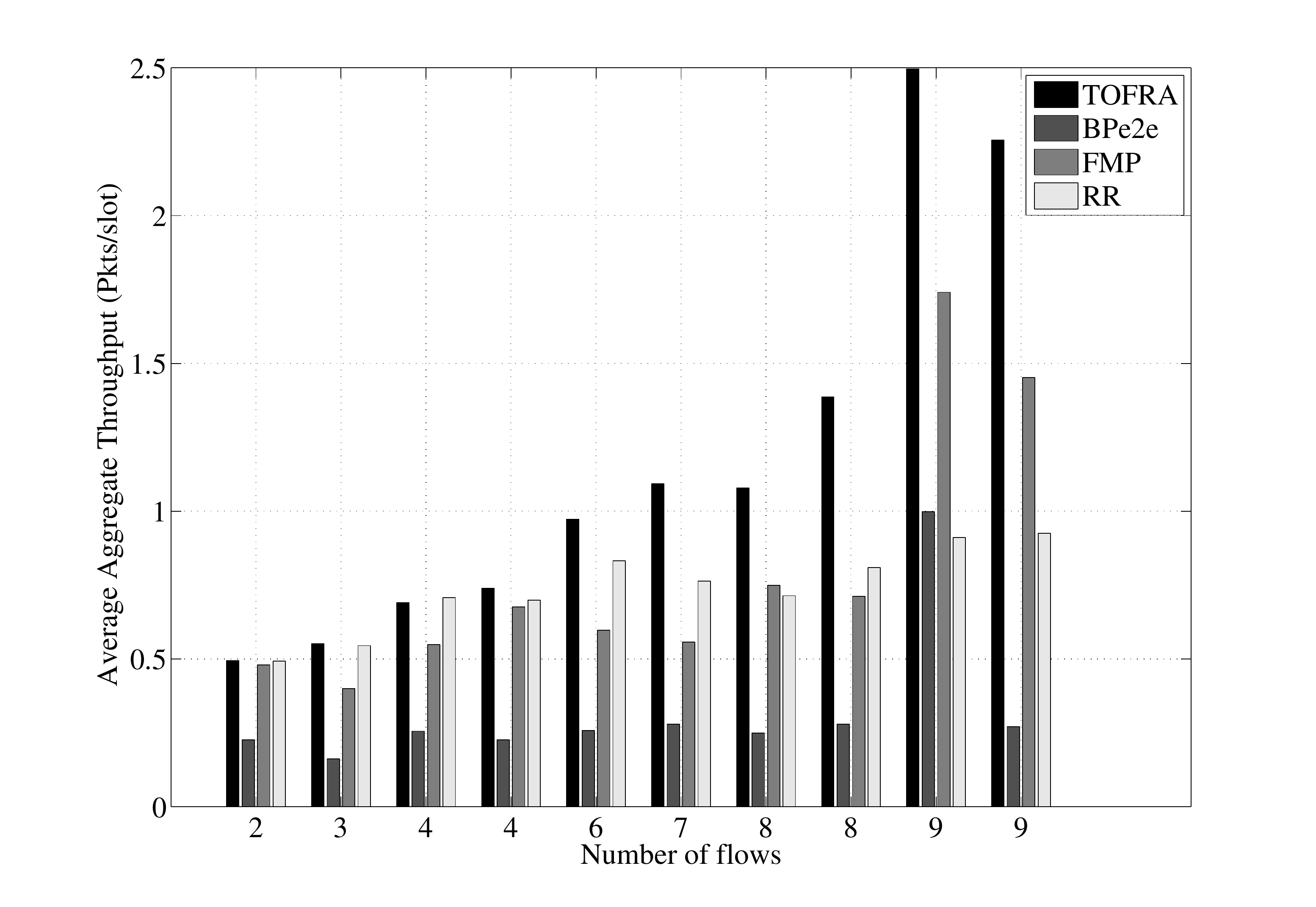}
\caption{Simulation results: AAT for TOFRA, FMP, BP, and RR for $\gamma$ $=$ $0.5$.}
\label{fig_comparison_g05}
\end{figure}

\begin{figure}
\centering
\includegraphics[scale=0.30]{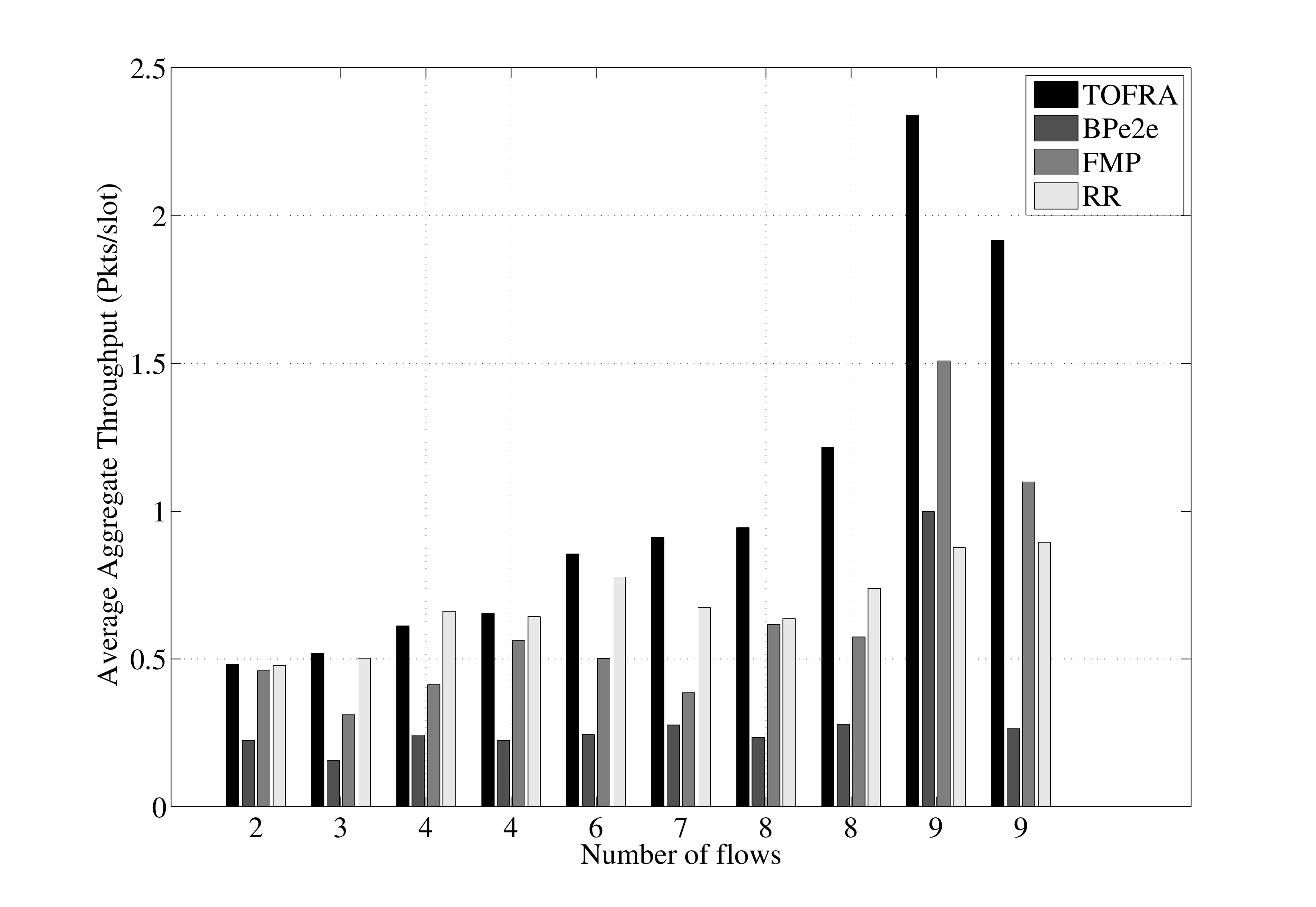}
\caption{Simulation results: AAT for TOFRA, FMP, BP, and RR for $\gamma$ $=$ $1.0$.}
\label{fig_comparison_g10}
\end{figure}

\begin{figure}
\centering
\includegraphics[scale=0.30]{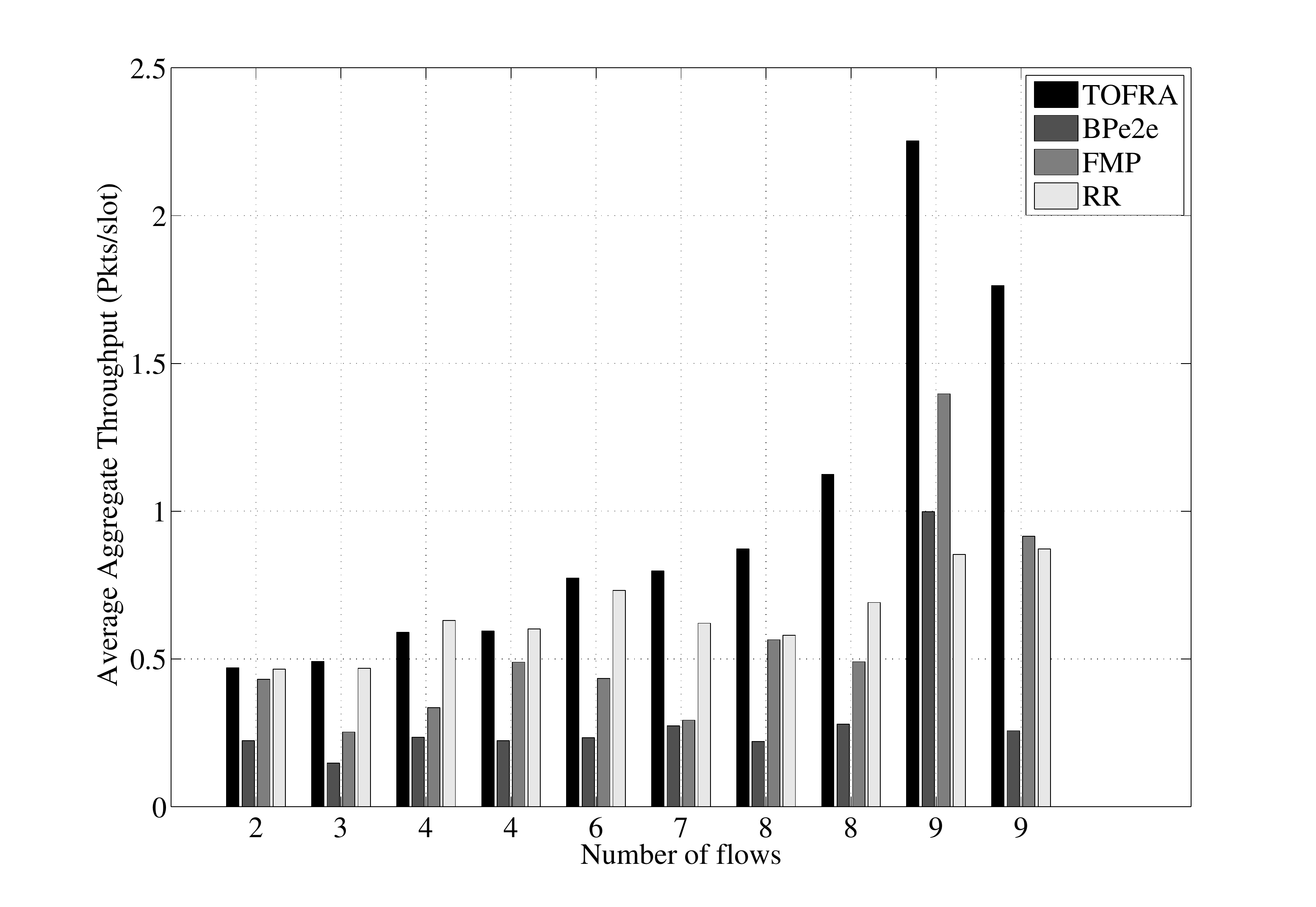}
\caption{Simulation results: AAT for TOFRA, FMP, BP, and RR for $\gamma$ $=$ $1.5$.}
\label{fig_comparison_g15}
\end{figure}

\begin{figure}
\centering
\includegraphics[scale=0.30]{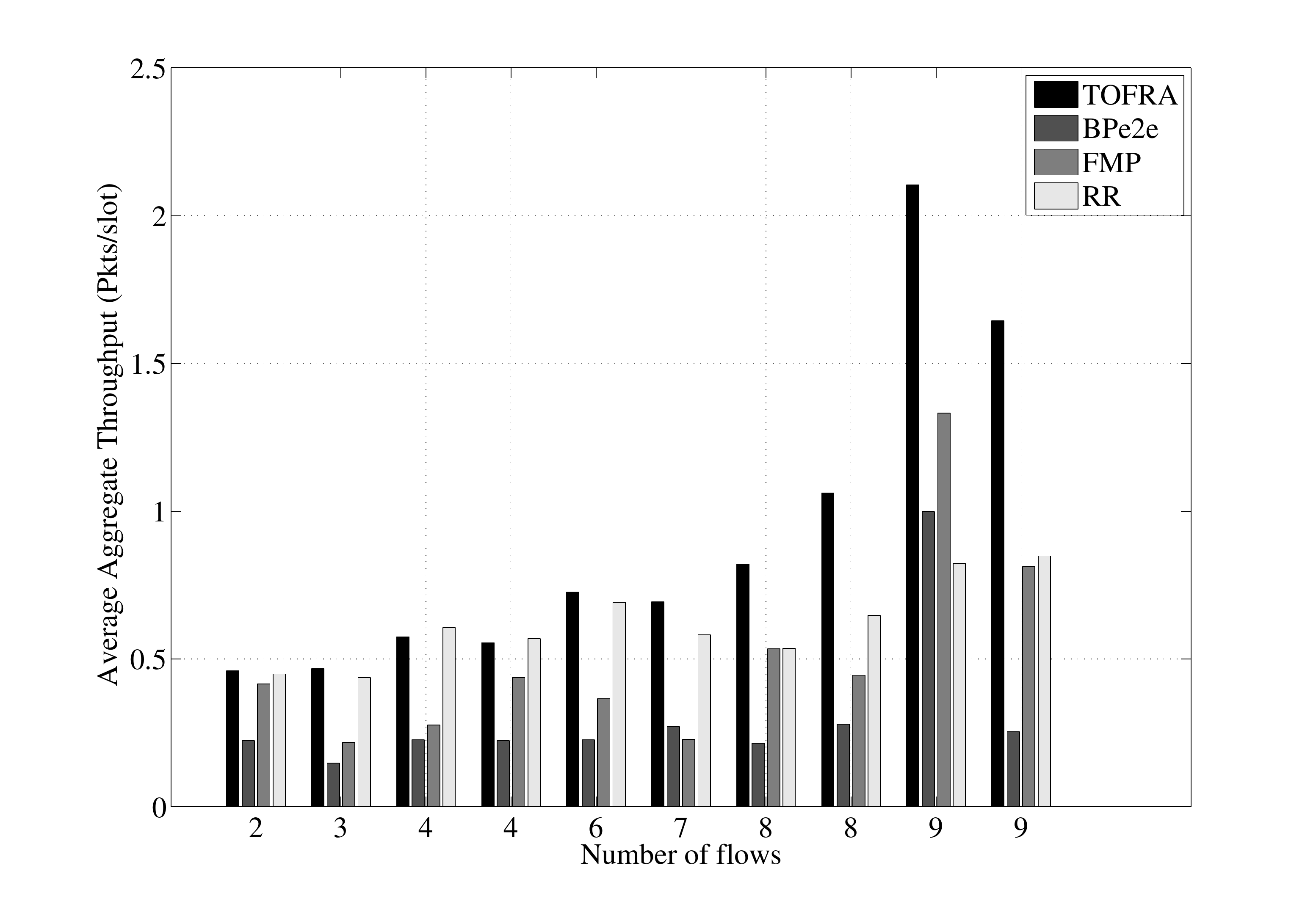}
\caption{Simulation results: AAT for TOFRA, FMP, BP, and RR for $\gamma$ $=$ $2.0$.}
\label{fig_comparison_g20}
\end{figure}

In the second part of the evaluation process, the proposed flow allocation scheme (TOFRA) is compared, in terms of AAT,
with three other flow allocation schemes, namely,
\textit{Best-Path (BP)}, \textit{Full MultiPath (FMP)}, and \textit{Round-Robin (RR)}.
BP employs a single path to the destination, which is the one that exhibits the highest end-to-end success probability
(defined in Section \ref{sec:mpath_mhop_analysis}) and estimates the flow that should be assigned on this path, by solving a single path version
of the flow allocation optimization problem.
FMP assigns a flow rate of one packet per slot on each path, while RR employs a different path each time slot.
For the evaluation process, we consider the simulated variant for each scheme, where the assumption of saturated queues is removed.

Figs.~\ref{fig_comparison_g05}-\ref{fig_comparison_g20}, collate the AAT achieved by all aforementioned schemes, for the ten random scenarios employed.
Each figure corresponds to one of the different SINR threshold values considered ($0.5$, $1.0$, $1.5$, and $2.0$).
As these figures show, the proposed flow allocation scheme (TOFRA)
achieves significantly higher ATT than FMP. The main reason for this is that, it takes into account
the effect of both intra- and inter-path interference on throughput. FMP on the other hand, assigns the maximum flow data rate on each path
(one packet per slot), disregarding the effect of interference. TOFRA achieves $47.3\%$, $63.7\%$, $78.9\%$, and $91.5\%$ higher AAT,
on average, over all ten scenarios, than FMP, for $\gamma =  0.5, 1.0, 1.5, 2.0 $, respectively.
The proposed scheme also outer-performs BP, for all traffic scenarios and $\gamma$ values. This is however expected, since
TOFRA exploits the diversity among the available paths and is able to aggregate resources from different paths on and interference-aware
manner. The average gain of TOFRA over BP is $293.7\%$, $256.4\%$, $2391.1\%$, and $222.1\%$, for the four $\gamma$ values considered.

As far as round robin (RR) scheme is concerned, the average gain of TOFRA over RR, in terms of AAT, is $50.5\%$, $43.7\%$, $41.7\%$,
and $39.1\%$, for $\gamma =  0.5, 1.0, 1.5, 2.0 $, respectively.
Comparing TOFRA with RR reveals the following trend: in scenarios where a low number of
flows is present ($\leq4$), the gain of TOFRA over RR is insignificant. Moreover, in specific scenarios, and especially when a larger $\gamma$ value
is employed, RR achieves slightly higher AAT than TOFRA. This is the case for scenario $2$ and all $\gamma$ values, and scenario $3$ and $\gamma$ values
$1.5$, and $2.0$, respectively. In scenarios with a larger number of flows, TOFRA outer-performs RR.
The advantage of RR over TOFRA is that, alternating among the available paths, on an iterative manner, it reduces both inter-path interference
and packet failures along each path, due to half-duplex node operation. However, round-robin based flow allocation is expected to exhibit poor
performance in two cases: firstly, in scenarios where a larger number of flows is present and thus, a larger number of paths is utilized. In a scenario
with $K$ flows, employing $K$ paths for example, each path will remain idle before being assigned another packet to forward, for $K-1$ slots.
Secondly, RR is expected to achieve significantly lower AAT than TOFRA in scenarios where there is a large degree of diversity among the
available paths. The reason for this is that, RR assigns packets on paths on a periodic manner, without adjusting flow rate based on
their quality.

\begin{figure}
\centering
\includegraphics[scale=0.30]{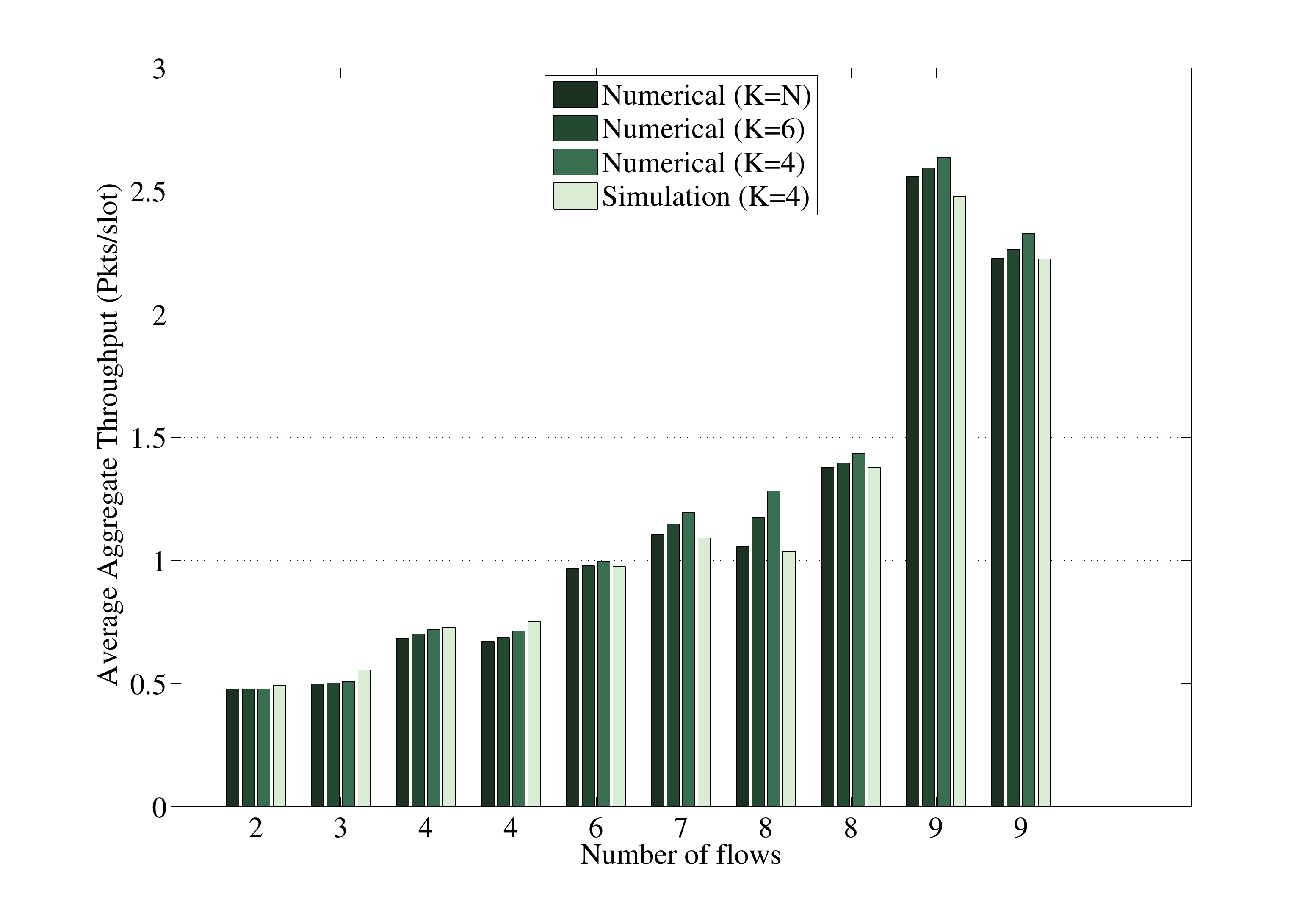}
\caption{Effect of number of dominant interferers on average aggregate throughput accuracy for $\gamma$ $=$ $0.5$.}
\label{fig:kdominant_g05}
\end{figure}
\begin{figure}
\centering
\includegraphics[scale=0.30]{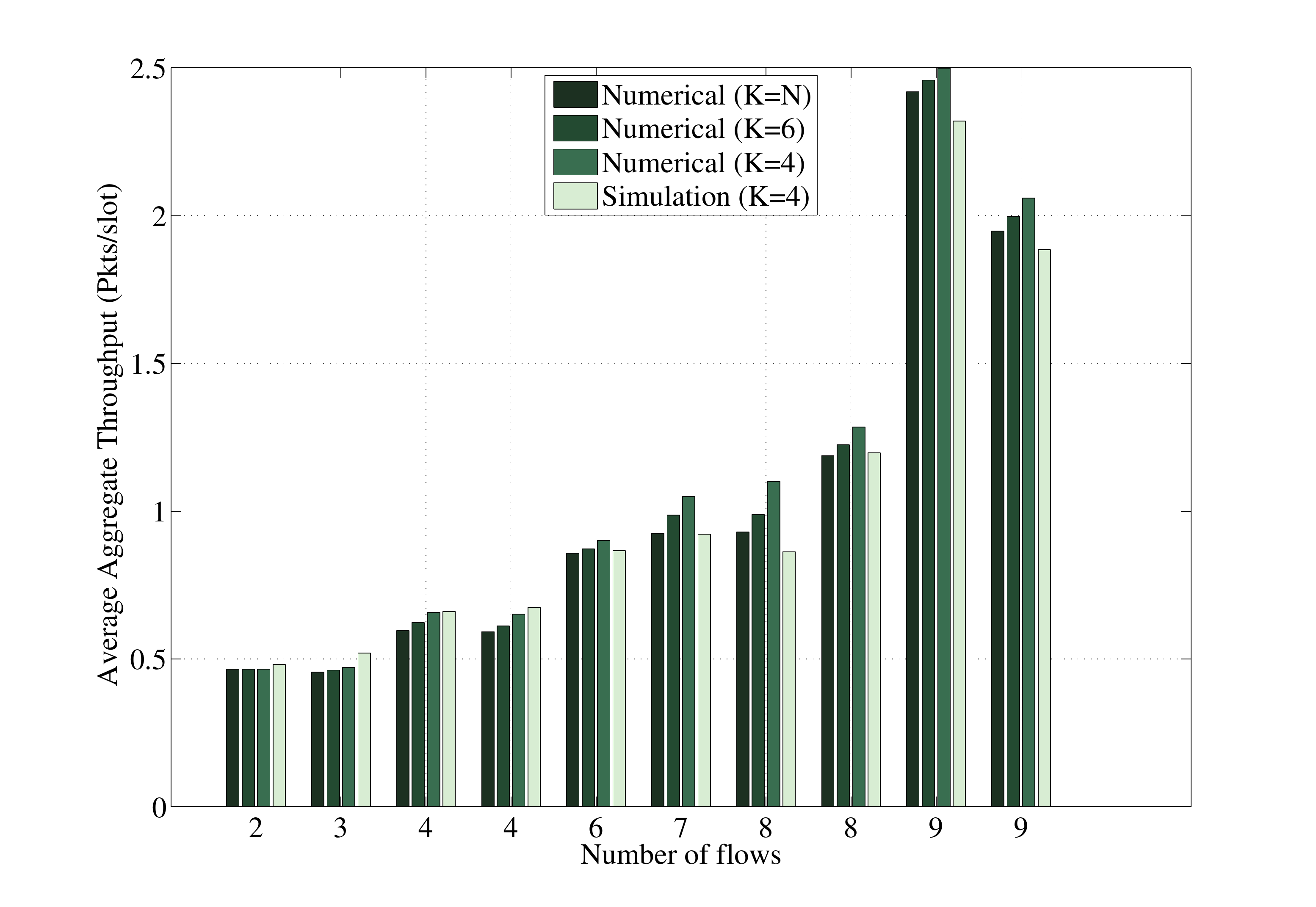}
\caption{Effect of number of dominant interferers on average aggregate throughput accuracy for $\gamma$ $=$ $1.0$.}
\label{fig:kdominant_g10}
\end{figure}
\begin{figure}
\centering
\includegraphics[scale=0.30]{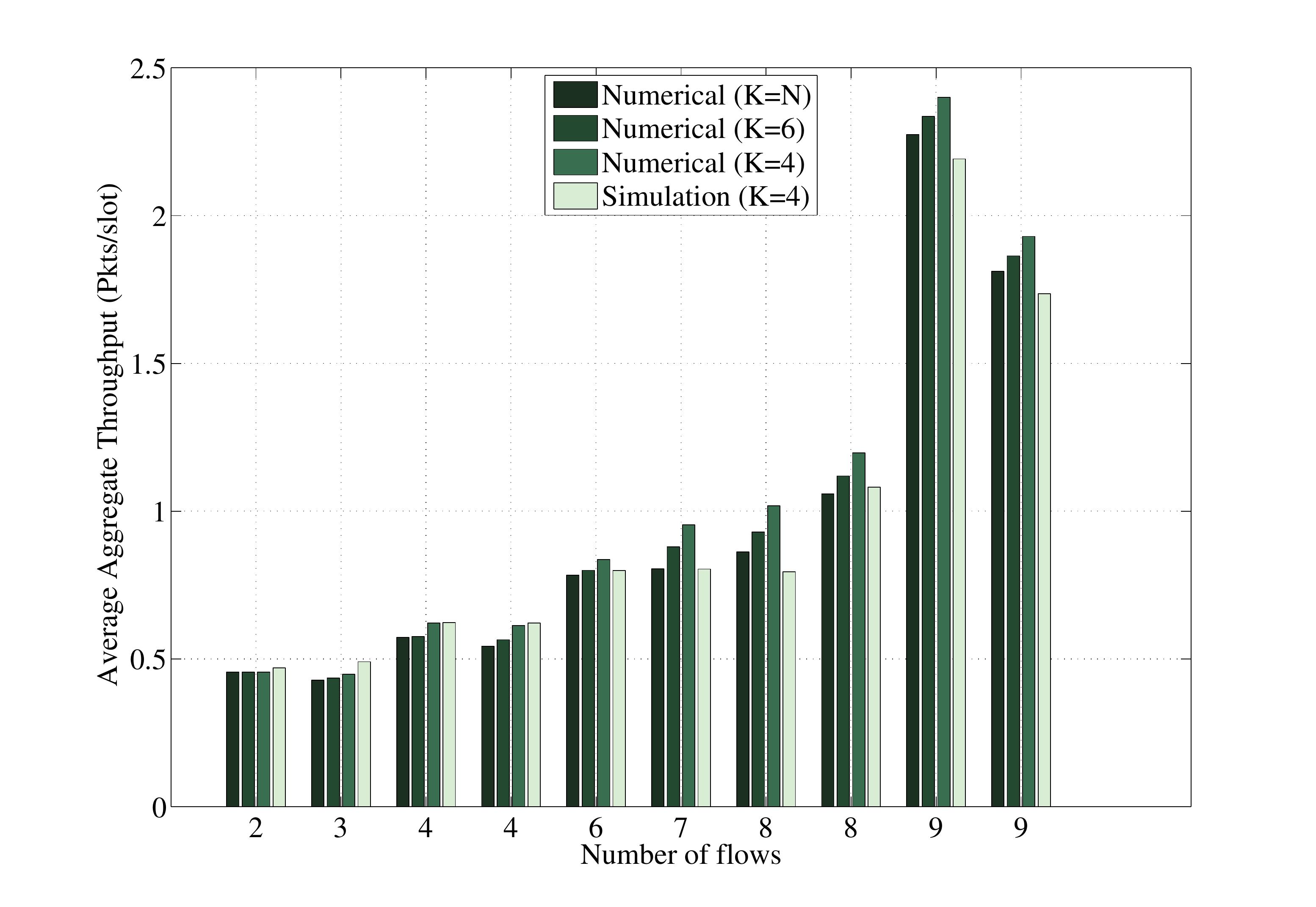}
\caption{Effect of number of dominant interferers on average aggregate throughput accuracy for $\gamma$ $=$ $1.5$.}
\label{fig:kdominant_g15}
\end{figure}
\begin{figure}
\centering
\includegraphics[scale=0.30]{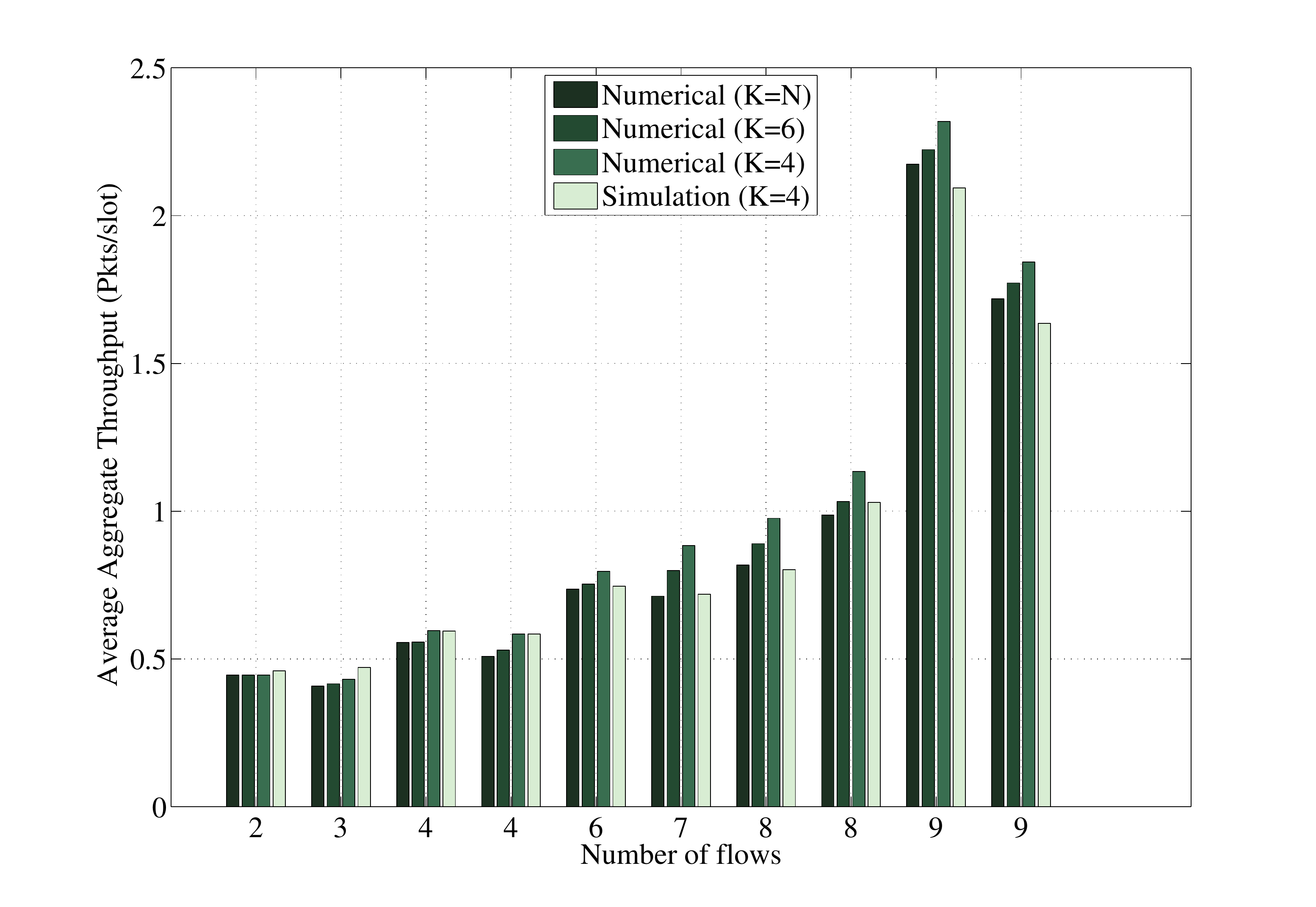}
\caption{Effect of number of dominant interferers on average aggregate throughput accuracy for $\gamma$ $=$ $2.0$.}
\label{fig:kdominant_g20}
\end{figure}

In the last part of the evaluation process, a variant of the proposed scheme is explored, where interference is approximated
by considering only the \textit{dominant} interfering nodes for each link. The goal is
to reduce the complexity of expressing the average aggregate throughput (AAT) achieved by all flows and consequently, of solving the flow allocation
optimization problem. As already described in Section \ref{sec:analysis}, the first step of the process for formulating flow allocation
as an optimization problem, is deriving the expression for a link's average throughput.
Instead of considering all possible interfering nodes for expressing the average throughput achieved over that link,
the interference imposed on it, is approximated by taking into account only the \textit{k dominant} ones.
The term \textit{dominant interfering nodes} refers to transmitters that contribute with the most significant amount of interference, on average,
to packet receptions over a specific link and thus, have the most significant effect on its success probability.
Approximating the interference imposed on a link in this way, under-estimates the actual interference experienced by that link
in the simulated scenarios, and thus, results in an over-estimation of the AAT.

The purpose of this part of the evaluation process is to explore the trade-off between, reduced complexity
in formulating flow allocation as an optimization problem and accuracy in capturing the average aggregate throughput observed
in the simulated scenarios.
For each wireless scenario and $\gamma$ value ($0.5$, $1.0$, $1.5$, and $2.0$), the flow allocation problem employed by the TOFRA scheme,
is formulated and solved through the simulated annealing technique, considering each time a different number of dominant interfering nodes
($K=2,...,6$). In this way, the proposed scheme estimates the rates that achieve maximum AAT, along with the corresponding AAT value,
for each wireless scenario, $\gamma$ value, and different number of interfering nodes. Numerical results concerning AAT that are estimated
on this way are presented in Figs.~\ref{fig:kdominant_g05}-\ref{fig:kdominant_g20}, with labels \textit{Numerical (K=N)},
\textit{Numerical (K=6)}, and \textit{Numerical (K=4)},
based on the number of dominant interfering nodes considered. Note that, label \textit{Numerical (K=N)}, indicates numerical results derived by the flow allocation
optimization problem, by considering all interfering nodes for each link.

When the number of interfering nodes considered, for expressing each link's average throughput is reduced, TOFRA overestimates
the maximum AAT that can be achieved, by all flows present in the wireless scenario considered. Comparing numerical results concerning
TOFRA's AAT, for $K=N$, and $K=6$, the average overestimation over all wireless scenarios is $2.6\%$, $3.0\%$, $3.6\%$, and $3.9\%$, for
$\gamma = 0.5, 1.0, 1.5,$ and $2.0$, respectively. The corresponding values, for the case where numerical results for $K=N$, and $K=4$ are compared,
are $5.7\%$, $7.7\%$, $9.4\%$, and $10.8\%$. As Figs.~\ref{fig:kdominant_g05}-\ref{fig:kdominant_g20} also show, this overestimation
becomes more significant for large $\gamma$ values, where the effect of interference on success probability becomes more acute.
These results show that, considering only a small number of dominant interfering nodes for each link,
results in TOFRA estimating an AAT value, that differs insignificantly from the one estimated when all interfering nodes are taken into account.

What is most interesting though, is to explore whether the AAT estimated through TOFRA's model, when considering only the $K$ dominant interfering nodes for each
link, differs significantly from the actual AAT observed in the simulated scenarios. It is also important to note that, while estimating the received
SINR for a specific packet, all active transmitters are taken into account for interference inference, implying that, the actual
interference experienced by a link in the simulated scenarios, is higher than the one considered by the TOFRA variant, where interference
for each link is approximated by considering the dominant interfering nodes only.

Observing Figs.~\ref{fig:kdominant_g05}-\ref{fig:kdominant_g20} shows that, for all $\gamma$ values and for most scenarios, TOFRA's AAT,
in the simulated scenarios, is lower than the one estimated by the analysis employed (flow allocation optimization problem). This is expected
however, since in the analysis, only a subset of all the interfering nodes (the dominant ones) are considering for expressing a specific
link's average throughput. To be more precise, TOFRA's AAT observed in the simulated scenarios, is lower than the corresponding numerical values
for $60\%$ of the wireless scenarios, for $\gamma = 0.5, 1.0, 1.5,$ and for $70\%$ of them for $\gamma =2.0$. It is also interesting to note that,
in some scenarios, the simulated AAT is higher than the one estimated by the flow allocation optimization problem. The reason for this, was also discussed
in the first part of the evaluation process, in the beginning of this section, and is related to the saturated queues assumption present in our analysis.
Even if only the dominant interfering nodes are considered for a specific link, it is assumed that these nodes will always have a packet available
for transmission in their queues. However, this is not always the case in the simulated scenarios and so, the actual interference experienced
by a link, from these dominant interferers, may be lower than the one estimated by our analysis. In this way, the effect of interference underestimation,
by considering only the dominant interfering nodes for each link, is counter-balanced. For each wireless scenario and $\gamma$ value, the absolute
value of the deviation between numerical and simulated AAT is estimated, for the case, where both of them are derived by considering only the
four dominant interfering nodes for each link. The average value of this deviation, over all wireless scenarios, is $6.8\%$, $8.5\%$, $9.6\%$, and
$9.6\%$, for $\gamma = 0.5, 1.0, 1.5,$ and $2.0$ respectively. These results show that, the gain of reduced complexity for expressing a link's average
throughput, comes at an insignificant cost in the accuracy with which simulated AAT for the proposed scheme is captured by the analysis employed.

\section{Conclusion}
\label{sec:conclusions}

This study, explores the issue of aggregate throughput optimal flow rate allocation, for wireless, multi-hop,
random access networks, with multi-packet reception capabilities.
Flows are forwarded over multiple node disjoint paths, experiencing both intra- and inter-path interference.
We propose a distributed scheme that formulates flow rate allocation as an optimization problem, aimed at maximizing
the average aggregate throughput of all flows, while also providing bounded packet delay guarantees.
The key feature of the suggested scheme is that, it employs a simple model for the average aggregate throughput, that
captures both intra- and inter-path interference through the SINR model.
A simple topology is employed to demonstrate the proposed scheme and also show that the corresponding optimization problem is non-convex.
We evaluate the suggested flow allocation scheme using Ns2 simulations of ten random wireless scenarios.
Collating numerical with simulations results reveals that, the suggested scheme accurately captures the average aggregate
throughput observed in the simulated scenarios, despite the simplifying assumptions adopted by our analysis.
Moreover, it achieves significantly higher average aggregate throughput than best-path, full multipath and a
round-robin based flow allocation scheme.
As part of the evaluation process, we also explore the trade-off between, reduced complexity in formulating flow allocation
as an optimization problem and the accuracy in estimating the average aggregate throughput observed in the simulation results.
A variant of the proposed scheme is explored, where interference for each link is approximated by considering the dominant
interfering nodes only for expressing its average throughput.
Simulation results show that, approximating the interference experienced by each link by considering only the dominant interfering nodes,
results in an insignificant deviation between numerical and simulation results, regarding AAT.

Part of our future work, is to address fairness issues too, apart from maximizing the aggregate throughput achieved by all flows.
We also plan to consider multiple transmission rates and relax the assumption of fixed transmission probability per relay,
by allowing a variable contention window.
In the present study we treat interference as noise. In future steps, we aim at adopting more sophisticated approaches
for interference handling, such as, successive interference cancellation and joint decoding \cite{b:Tse, b:PloumidTOFRASIC}.
Finally, we aim at exploring the performance of the suggested flow rate allocation scheme under the assumption of bursty packet losses.

\ifCLASSOPTIONcaptionsoff
  \newpage
\fi

\bibliographystyle{ieeetr}
\bibliography{bibliography_may_2015}

\end{document}